\documentclass[twocolumn,tighten,times]{aastex63}
\usepackage{sidecap}
\usepackage{amsmath,amstext}
\usepackage{mathtools}
\usepackage{afterpage}
\usepackage{lipsum}
\usepackage{capt-of}
\usepackage{relsize}
\usepackage{newtxtext,newtxmath}
\usepackage[all]{hypcap} 
\usepackage{breqn}
\usepackage{verbatim}
\usepackage{graphicx}
\usepackage{footnote}
\usepackage{float}
\usepackage{amsmath}
\usepackage{enumitem}
\usepackage{amssymb}
\usepackage{array}
\usepackage{gensymb}
\usepackage[graphicx]{realboxes}
\usepackage[version=4]{mhchem}
\usepackage{graphicx,epsf}
\usepackage{subfigure}
\usepackage{natbib}

\usepackage{rotating}
\usepackage{nicefrac}
\usepackage{xtab} 
\usepackage{soul}

\newcommand\aastex{AAS\TeX}

\usepackage{savesym}
\savesymbol{tablenum}
\restoresymbol{SIX}{tablenum}

\newcommand{\ztf}{\textit{ZTF}}
\newcommand{\decam}{\textit{DECam}}
\newcommand{\atlas}{\textit{ATLAS}}
\newcommand{\tsf}{\text{SN~2019tsf}}
\newcommand{\hls}{\text{iPTF14hls}}

\newcommand{\Msun}{{\rm M}_\odot}

\newcommand{\niVI}{${}^{56}\textrm{Ni}$}
\newcommand{\coVI}{${}^{56}\textrm{Co}$}

\newcommand{\feVI}{${}^{56}\textrm{Fe}$}
\newcommand{\be}{\begin{equation}}
\newcommand{\ee}{\end{equation}}
\newcommand{\CHE}{CHE Israel Excellence Fellowship}
\newcommand{\NSF}{National Science Foundation Graduate Research Fellow}

\graphicspath{{./}{figures/}}

\shorttitle{3-peaks \tsf{}}
\shortauthors{Zenati et al.}

\begin{document}

\title{Template \aastex Article with Examples: 
v1\footnote{Released on April, 26th, 2021}}

\title{\tsf: Evidence for Extended Hydrogen-Poor CSM in the Three-Peaked Light Curve of Stripped Envelope of a Type Ib Supernova}

\correspondingauthor{Yossef Zenati}
\email{yzenati1@jhu.edu}

\author[0000-0002-0632-8897]{Yossef Zenati}
\altaffiliation{\CHE}
\affil{Physics and Astronomy Department, Johns Hopkins University, Baltimore, MD 21218, USA}
\affil{Space Telescope Science Institute, 3700 San Martin Dr., Baltimore, MD 21218, USA}
\affil{Astrophysics Research Center of the Open University (ARCO), The Open University of Israel, Ra’anana 4353701, Israel}

\author[0000-0001-5233-6989]{Qinan Wang}
\affil{Physics and Astronomy Department, Johns Hopkins University, Baltimore, MD 21218, USA}

\author[0000-0002-4674-0704]{Alexey Bobrick}
\affil{Technion - Israel Institute of Technology, Physics department, Haifa Israel 3200002}

\author[0000-0003-4587-2366]{Lindsay DeMarchi}
\affil{Center for Interdisciplinary Exploration and Research in Astrophysics (CIERA) and Department of Physics and Astronomy, Northwestern University, Evanston, IL 60208}

\author[0000-0002-6012-2136]{Hila Glanz}
\affil{Technion - Israel Institute of Technology, Physics department, Haifa Israel 3200002}

\author[0000-0002-2728-0132]{Mor Rozner}
\affil{Technion - Israel Institute of Technology, Physics department, Haifa Israel 3200002}

\author[0000-0001-5754-4007]{Jacob E. Jencson}
\affiliation{IPAC, California Institute of Technology, 1200 E. California Boulevard, Pasadena, CA 91125, USA}
\affil{Physics and Astronomy Department, Johns Hopkins University, Baltimore, MD 21218, USA}
\affil{Space Telescope Science Institute, 3700 San Martin Dr., Baltimore, MD 21218, USA}

\author[0000-0002-4410-5387]{Armin Rest}
\affil{Space Telescope Science Institute, 3700 San Martin Dr., Baltimore, MD 21218, USA}
\affil{Physics and Astronomy Department, Johns Hopkins University, Baltimore, MD 21218, USA}

\author[0000-0002-4670-7509]{Brian D.\ Metzger}
\affil{Department of Physics and Columbia Astrophysics Laboratory, Columbia University, Pupin Hall, New York, NY 10027, USA}
\affil{Center for Computational Astrophysics, Flatiron Institute, 162 5th Ave, New York, NY 10010, USA} 

\author[0000-0003-4768-7586]{Raffaella Margutti}
\affil{Department of Astronomy and Astrophysics, University of California, Berkeley, CA 94720, USA}

\author[0000-0001-6395-6702]{Sebastian Gomez}
\affil{Space Telescope Science Institute, 3700 San Martin Dr., Baltimore, MD 21218, USA}

\author[0000-0001-5510-2424]{Nathan Smith}
\affil{Steward Observatory, University of Arizona, 933 N. Cherry Avenue, Tucson, AZ 85721, USA}

\author[0000-0002-2998-7940]{Silvia Toonen}
\affil{Anton Pannekoek Institute for Astronomy, University of Amsterdam, 1090 GE Amsterdam, The Netherlands}

\author[0000-0002-7735-5796]{Joe S.\ Bright}
\affil{Department of Astronomy, University of California, Berkeley, CA 94720-3411, USA}
\affil{Astrophysics, Department of Physics, University of Oxford, Keble Road, Oxford OX1 3RH, UK}

\author[0000-0002-4410-5387]{Colin Norman}
\affil{Physics and Astronomy Department, Johns Hopkins University, Baltimore, MD 21218, USA}
\affil{Space Telescope Science Institute, 3700 San Martin Dr., Baltimore, MD 21218, USA}

\author[0000-0002-2445-5275]{Ryan J.\ Foley}
\affil{Department of Astronomy and Astrophysics, University of California, Santa Cruz, CA 95064, USA}

\author[0000-0001-6022-0484]{Alexander Gagliano}
\altaffiliation{\NSF}
\affil{Department of Astronomy, University of Illinois at Urbana-Champaign, 1002 W. Green St, IL 61801, USA}
\affil{Center for Astrophysical Surveys, National Center for Supercomputing Applications, Urbana, IL 61801, USA}
\affil{Center for Astrophysical Surveys, Urbana, IL 61801, USA}

\author[0000-0002-2995-7717]{Julian~H.~Krolik}
\affiliation{Physics and Astronomy Department, Johns Hopkins University, Baltimore, MD 21218, USA}

\author[0000-0002-8229-1731]{Stephen J.\ Smartt}
\affil{Astrophysics Research Centre, School of Mathematics and Physics, Queens University Belfast, Belfast BT7 1NN, UK}

\author[0000-0002-5814-4061]{Ashley V.\ Villar}
\affil{Department of Astronomy and Astrophysics, Pennsylvania State University, 525 Davey Laboratory, University Park, PA 16802, USA}
\affil{Institute for Computational \& Data Sciences, The Pennsylvania State University, University Park, PA, USA}
\affil{Institute for Gravitation and the Cosmos, The Pennsylvania State University, University Park, PA 16802, USA}

\author[0000-0001-6022-0484]{Gautham Narayan}
\affil{Department of Astronomy, University of Illinois at Urbana-Champaign, 1002 W. Green St, IL 61801, USA}
\affil{Center for Astrophysical Surveys, National Center for Supercomputing Applications, Urbana, IL 61801, USA}
\affil{Center for Astrophysical Surveys, Urbana, IL 61801, USA}

\author[:0000-0003-2238-1572]{Ori Fox}
\affil{Space Telescope Science Institute, 3700 San Martin Dr., Baltimore, MD 21218, USA}

\author[0000-0002-4449-9152]{Katie Auchettl}
\affil{Department of Astronomy and Astrophysics, University of California, Santa Cruz, CA 95064, USA}
\affil{School of Physics, The University of Melbourne, VIC 3010, Australia}
\affil{ARC Centre of Excellence for All Sky Astrophysics in 3 Dimensions (ASTRO 3D), Australia}

\author{Daniel Brethauer}
\affil{Department of Astronomy and Astrophysics, University of California, Berkeley, CA 94720, USA}

\author{Alejandro Clocchiatti}
\affil{Instituto de Astrof\'{\i}sica, Pontificia Universidad Cat\'olica,
Vicu\~na Mackenna 4860, 7820436 Santiago, Chile}
\affil{Millennium Institute of Astrophysics, Nuncio Monse\~nor ˜ S\'otero Sanz 100, Of. 104, Providencia, 7500000 Santiago, Chile}

\author[0000-0002-8436-5431]{Sophie V. Coelln}
\affil{Physics and Astronomy Department, Johns Hopkins University, Baltimore, MD 21218, USA}

\author[0000-0001-5126-6237]{Deanne L. Coppejans}
\affil{Department of Physics, University of Warwick, Coventry CV4 7AL, UK}

\author[0000-0001-9494-179X]{Georgios Dimitriadis}
\affil{School of Physics, Trinity College Dublin, The University of Dublin, Dublin 2, Ireland}

\author{Andris Dorozsmai}
\affil{Institute of Gravitational Wave Astronomy and School of Physics and Astronomy, University of Birmingham, Edgbaston, Birmingham B15 2TT, United Kingdom}
\affil{Anton Pannekoek Institute for Astronomy, University of Amsterdam, 1090 GE Amsterdam, The Netherlands}

\author[0000-0001-7081-0082]{Maria Drout}
\affil{David A. Dunlap Department of Astronomy and Astrophysics, University of Toronto, 50 St. George Street, Toronto, Ontario, M5S 3H4, Canada}
\affil{Observatories of the Carnegie Institution for Science, 813 Santa Barbara Street, Pasadena, CA 91101, USA}

\author[0000-0002-4674-0704]{Wynn Jacobson-Galan}
\affil{Department of Astronomy and Astrophysics, University of California, Berkeley, CA 94720, USA}

\author[0000-0002-9902-6803]{Bore Gao}
\affil{Physics and Astronomy Department, Johns Hopkins University, Baltimore, MD 21218, USA}

\author[0000-0003-1724-2885]{Ryan Ridden-Harper}
\affil{School of Physical and Chemical Sciences | Te Kura Matū, University of Canterbury, Private Bag 4800, Christchurch 8140, New Zealand}

\author[0000-0002-5740-7747]{Charles Donald Kilpatrick}
\affil{Center for Interdisciplinary Exploration and Research in Astrophysics (CIERA) and Department of Physics and Astronomy, Northwestern University, Evanston, IL 60208}

\author[0000-0003-1792-2338]{Tanmoy Laskar}
\affil{Department of Astrophysics/IMAPP, Radboud University, PO Box 9010, 6500 GL, The Netherlands}

\author[0000-0002-4513-3849]{David Matthews}
\affil{Department of Astronomy and Astrophysics, University of California, Berkeley, CA 94720, USA}

\author[0000-0002-3825-0553]{Sofia Rest}
\affil{Physics and Astronomy Department, Johns Hopkins University, Baltimore, MD 21218, USA}

\author[0000-0001-9535-3199]{Ken W. Smith}
\affil{Astrophysics Research Centre, School of Mathematics and Physics, Queen's University Belfast, Belfast, BT7 1NN, UK}

\author{Candice McKenzie Stauffer}
\affil{Center for Interdisciplinary Exploration and Research in Astrophysics (CIERA) and Department of Physics and Astronomy, Northwestern University, Evanston, IL 60208}

\author[0000-0002-3019-4577]{Michael C. Stroh}
\affil{Center for Interdisciplinary Exploration and Research in Astrophysics (CIERA) and Department of Physics and Astronomy, Northwestern University, Evanston, IL 60208}

\author[0000-0002-7756-4440]{Louis-Gregory Strolger}
\affil{Space Telescope Science Institute, 3700 San Martin Dr., Baltimore, MD 21218, USA}

\author[0000-0003-0794-5982]{Giacomo Terreran}
\affil{Center for Interdisciplinary Exploration and Research in Astrophysics (CIERA) and Department of Physics and Astronomy, Northwestern University, Evanston, IL 60208}

\author[0000-0002-2361-7201]{Justin D. R. Pierel}
\affil{Space Telescope Science Institute, 3700 San Martin Dr., Baltimore, MD 21218, USA}

\author[0000-0001-6806-0673]{Anthony L.\ Piro}
\affil{The Observatories of the Carnegie Institution for Science, 813 Santa Barbara St., Pasadena, CA 91101, USA}


\begin{abstract}

We present multi-band \atlas{} and \ztf{} photometry for \tsf{}, a Type Ib stripped-envelope supernova (SESN). The slow spectral evolution could be associated with an uncommon explosion mechanism specific to this SN. Possible explanations include fallback accretion onto a compact remnant or a long-lived central engine, both of which could provide extended energy injection responsible for the late-time re-brightening and unusual spectral features.

The re-brightening observations represent the latest photometric measurements of a multi-peaked Type Ib SN. As late-time photometry and spectroscopy suggest no hydrogen, the potential circumstellar material (CSM) must be H-poor. The absence of a nebular phase and the lack of narrow emission lines in the late-time spectra ($>$ 142 days) of the SNe suggest that any CSM interaction is likely asymmetric and enveloped by the SN ejecta.

However, an extended CSM structure is evident through a follow-up radio campaign with the Karl G. Jansky Very Large Array (VLA), indicating a source of bright optically thick radio emission at late times, which is highly unusual among H-poor SESNe. We attribute this phenomenology to an interaction of the supernova ejecta with asymmetric CSM, potentially disk-like, and we present several models that may explain the origin of this rare Type Ib supernova. We propose a warped disk model in which a tertiary companion—commonly present around massive stars-perturbs the progenitor's CSM, producing density enhancements that may explain the observed multi-peaked \tsf{} light curve. This \tsf{} is a unique SN Type Ib among the recently discovered class of SNe that undergo mass transfer at the moment of explosion.

\end{abstract}


\section{Introduction} \label{sec:intro}

There is significant observational diversity among stripped-envelope supernovae (SESNe), systems defined by the absence of H and He in their optical spectra. 
Following the standard SN classification scheme 
developed during the last three decades, these events can be H-poor and He-rich (SN~Ib); H- and He-poor (SN~Ic); or somewhere in between, with early H-$\alpha$ lines that fade with time (IIb SNe, \citealt{Filippenko+93, Woosley+95, Dessart+12, Yoon15, PrenticeAndMazzali17, GalYam2017, Prentice+19}). 

In their light curves, the typical time scale for the first peak of an SN Ib---believed to be powered by the radioactive decay of \niVI---is $\sim 20$~--~$25$ days \citep{Dessart+11, Taddia+15}. The light curves of SNe Ic, whose progenitors lack both their hydrogen and their helium envelopes, evolve like thermonuclear 
SNe~Ia, but are $\rm \sim 1\,{\rm mag}$ fainter at peak. These 
two subclasses are more common than SNe IIb; SNe Ib/c comprise $\sim 19\%$ of all SNe and $\sim 26\%$ of all core-collapse SNe (CCSNe; \citealp{Filippenko+93,Filippenko1997,Smith+11}, relative to $\sim 5$-$10\%$ of SNe II for SNe~IIb (e.g. \citealt{arcavi11,smith11,Claeys+11,Sana+12,GalYam2017,Fang+22}). SNe~IIb light curves may exhibit early bumps from the interaction of the SN shock or slower-moving ejecta with surrounding circumstellar material (CSM). Suppose the interaction of the shock wave with the CSM is particularly strong. In that case, the SN may also be classified as an SN Ibn (e.g., SN~2006jc, SN~2014av) due to the narrow spectral features of He, in analogy to the strongly interacting and H-rich SNe~IIn.
The light curves of these events are distinct among SESNe, with peaks near $\sim -19$ mag and a rapid decline $\rm \sim 0.1 mag /day$ \citep{Foley+07, smith06jc, Pastorello+16, SmithHB17, Hosseinzadeh+19}. 

More recently, \cite{Galyam+22} introduced the SN Icn class 
refer to objects that lack hydrogen and helium but show strong, narrow emission lines of carbon or oxygen (e.g., \citealt{Pellegrino22, Perley22}). Finally, Type I superluminous supernovae (SLSN-I) are a class of stripped-envelope core-collapse SNe characterized by their blue spectra and high luminosity, peaking at magnitudes brighter than $\rm \sim -19.8\ mag$ \citep{Chomiuk+11, Quimby+11, Villar+18, Gomez+21}.

Because the photometric and spectroscopic signatures of these systems are directly linked to the behavior of their progenitor systems in the pre-explosion phase, a thorough understanding of the diversity of SESN observations is paramount to uncovering the dynamic ecosystem of pathways that lead to stellar death. This diversity cannot be fully understood by the underlying \niVI{}\ mass synthesized in the explosion; for example, SNe~Ib/c and SNe~II both produce $\lesssim 0.03-0.1\Msun$ of \niVI\ \citep{Anderson19}, yet exhibit dramatically different photometric and spectroscopic evolution. Moreover, radioactive \niVI{} may not be the sole heating source for these enigmatic events; \cite{Ertl+19,Sollerman22} find that the amount of \niVI \ in SESN progenitor models are inadequate to give rise to the peak luminosities of $\sim$ half of ordinary Type Ib/c. 

Despite the numerous questions that remain in the effort to link SESN to their pre-explosion counterparts, growing samples have constrained the parameter space of viable explosion mechanisms for specific classes. Generally, two main channels are thought to give rise to SNe Ib/c. One channel is the mass loss from single massive stars via strong stellar winds:  
In this case, the progenitor is a Wolf-Rayet (WR) star (\citealp{Woosley+93, Georgy+09, Tramper+15, Dessart+20}, further details about the connection between SN Ib, Ic, and the WR in its different stages as, WNL, WNE, WC, and WO stars can be found in \citealt{Georgy+09}).
The second primary channel is binary interaction, which would require a close binary that ends its stellar evolution as a pair of young, massive stars \citep{YoonWoosley10,smith11, Sana+12, BenAmi+14, Dessart+15, Rimoldi+16, Janssens+21}.

Recent estimates suggest that mass-loss via stellar winds is generally insufficient to produce most SESN progenitors, with the exception of the most massive stars. Binary mass stripping has thus been proposed as an essential mechanism to explain the observed diversity of SESN \citep{smith14}. Most recently, \citet{fox22} discovered that the first surviving companion to a Type Ib/c in the case of SN 2013ge could be explained by binary models that tend to predict OB-type stars. Other more exotic scenarios have also been proposed for SLSNe-I; one theory consists of the collapsar model of a rapidly rotating star whose hydrogen envelope has been stripped during the pre-explosion time, which  is unlikely to explain the majority of SN Ib/c's
\citep{MacFadyenWoosley99,Woosley&Bloom06,Nicholl+17,Nagataki+18,Zenati+20,Lee+22}.

Multi-wavelength follow-up is uniquely valuable for understanding the explosion mechanisms of SESNe. Broad-lined SNe~Ic (SNe~Ic-BL), a subclass of SNe Ic with high expansion velocities and low host galaxy metallicities, remain the only supernova class to be unambiguously associated with long-duration gamma-ray bursts (LGRBs), whose gamma-ray emission lasts longer than $\sim$2 seconds \citep{Modjaz+19}. Since LGRBs naturally arise in the collapsar model for SESNe, the absence of LGRBs or the associated afterglows in most SNe Ib/c disfavors the collapsar theory for explaining these SNe \citep{Corsi+11_Ibc, Corsi+12}. \citet{iwamoto99} showed that the spectra of highly energetic events ($E_k\sim10^{52}$ ${\rm erg}$) are also accompanied by X-ray flashes (XRF), which further suggests that only that class of SNe is associated with LGRBs.

When obtaining multi-wavelength follow-up, full-phase coverage of the event is the key to providing a comprehensive picture of an explosion. Moderate cadence data can reveal distinct stages of an explosion, including its first emission in SNe~Ia (e.g., \citealt{qinanetal21}) and shock-breakout in Type-II SNe \citep{bersten18}. SESNe can show two prominent peaks in their optical light curves \citep{Roy+16, Gomez+19}: a burst of emission after the initial explosion, known as the shock cooling light curve \citep{arcavi+17a,GalYam2017}, and the radioactive decay of {\niVI} to {\coVI} and then {\coVI} to {\feVI} that powers the emission over the bulk of its lifetime and in late-time the rebrightening occurs due to CSM interaction.

In the last decade, transient surveys of increased sensitivity and depth have discovered SESNe with unprecedented photometric behavior. This has included the ultra-bright SLSNe-I and the ultra-rapid Fast Blue Optical Transients (FBOTs), the latter having a rise time to peak of less than $10$~days and a rapid fast decreasing exponential decline lasting $\sim 30$~days \citep{Arcavi+16}. Recently, \citet{Metzger2022} discussed a peculiar and extremely rare luminous FBOT subclass (LFBOTs). There are indications that these unique phenomena are realizations along a spectrum of non-standard interactions between SN ejecta and the surrounding CSM \citep{Kasen2017, LeungFullerNomoto21}. Because CSM is swept up as the SN shock and subsequent ejecta expand, these unusual light curve signatures offer an insightful window into the environments of SN progenitors at the moment of stellar death.

Here, we present optical and radio observations of \tsf{}, a SESN event with a remarkably unique light curve evolution. \tsf{} was initially classified as a SN~Ib by \citet{Sollerman+20}, and two distinct peaks were observed in the \ztf{} light curve: an initial peak (unfortunately at the beginning of observations) and a late-phase peak observed 90 days later. However, a {\it third} peak was uniquely observed by \atlas{} due to the extended coverage of the object, as shown later in this paper.

This evolution is distinct from any previously observed SNe Ib/c. Moreover, the spectra of \tsf{} have no discernible narrow-line hydrogen signatures that are typically the hallmark of CSM interactions, similar to \hls{} and SN~2020faa \citep{Arcavi14hls, Sollerman+19, Yang+212020faaSN}. Optical observations of \tsf{} extend over 430 days and reveal at least three well-resolved peaks until 180 days, during which the luminosity varies by as much as 50\%. Instead of lines of hydrogen formed by the explosion or the CSM interaction, spectroscopic follow-up of \tsf{} showed only weak hydrogen lines and a small fraction of helium. The closest known SN that had multiple bumps in the light curve similar to \tsf{} is \hls{} \citep{Arcavi14hls, Andrews&Smith18}.

Spectroscopically, \hls{} was also slowly evolving without unusual signatures such as narrow lines. However, narrow and intermediate-width lines from CSM interaction eventually appeared in the spectrum at very late times \citep{Arcavi14hls, Sollerman+19, Woosley18}. As inferred from observations, the CSM shells responsible for the observed peaks were likely ejected decades before the explosion. However, the CSM interaction might be strongly asymmetric, affecting the observed signatures \citep{Andrews&Smith18, Smith+15}. \hls~is slower than the typical Type IIP SNe and transitions into a SN-IIn at later phases. The SN showed an absorbing CSM with a relatively fast outflow of $\rm v_{CSM} \simeq 6000\ km s^{-1}$. Subsequently, \citet{Arcavi14hls} inferred that the explosion model requires the kinetic energy of the absorbing gas to be $\sim 10^{52}\,\rm erg$. SN~2020faa \citep{Yang+212020faaSN} is another supernova that displayed multiple peaks in its light curve, similar to SN \hls{}. SN~2020faa shows similarities to \hls{} in the first six months. However, there are differences, such as the absence of hydrogen and helium in \hls{} spectra. These differences in the spectral features suggest variations in the progenitor stars and explosion mechanisms. \tsf{} is mentioned as an example of a rare class of supernovae with multiple peaks. Unlike \hls{}, \tsf{} lacks hydrogen and helium in its spectra. Additionally, \tsf{} does not transition to the nebular phase even after 100 days after the peak, suggesting a prolonged evolution. The nature of its progenitor and the factors driving its unique behavior remain uncertain. Understanding these rare supernova events and their progenitors is challenging due to their complex and diverse nature. The presence or absence of hydrogen and helium lines, the timing of the peaks, and the characteristics of the CSM interaction provide valuable insights into the explosion mechanisms and the progenitor systems. Further observational and theoretical investigations and detailed modeling and analysis are necessary to unravel the complexities and shed light on the underlying physical processes associated with these exceptional supernovae. 
We further discuss \hls{} in Section \ref{subsec:SN2014hls} and \ref{sec:discussion}.

In this paper, we first present \atlas{} and late-phase \decam{} observations of \tsf{} and suggest theoretical scenarios to explain these light curves and spectra out to $\sim$~400\, days after the first detection. Moreover, we present the first epoch of radio observations of the SE \tsf{}. This epoch of radio observations uniquely allows us to build up a picture of the underlying physics of this phenomenon.

In \S\ref{sec:obs}, we present the optical observations and the data reduction of \tsf{}. In \S\ref{sec:radio_obs}, we present the Karl G. Jansky Very Large Array (VLA) observations and constrain the physical properties of the radio-emitting region with synchrotron modeling of the emission. In \S\ref{sec:lightcurve}, we construct the multi-band optical lightcurve for the SN.

In \S\ref{sec:lightcurve}, model the stellar evolution of \tsf{} and the light curve evolution and derive physical properties of the radioactive decay-powered explosion, magnetar, warped disk scenario, and the ejecta running in the disk and the interaction between SN ejecta and the CSM. In \S\ref{sec:discussion}, we discuss how \tsf{} compares to other late-time SN light curves and how these new observations constrain the SN progenitor system. In \S\ref{sec:conclusion}, we summarize our study with recommendations for future follow-up of the SESN-like \tsf{}.

In this paper, observed times are reported in UT. 
We adopt the AB magnitude system, unless where noted, and a flat $\Lambda$CDM cosmological model with $H_0 = 73$ km s$^{-1}$ Mpc$^{-1}$ \citep{riess2016, riess2018}. As discussed in \cite{Sollerman+20}, the host galaxy of \tsf{}, NGC 3541, has a well-established redshift of $z = 0.021$ and we adopt a distance of 83.9~Mpc (distance modulus $\mu = 34.64 \pm 0.54$\,mag; \citealp{}). A summary of the basic properties of the host galaxy and SN\,2019tsf is provided in Table~\ref{tbl:params}.

\begin{table}[t!]
\begin{center}
\caption{Main parameters of \tsf{} and its host galaxy \label{tbl:params}}
\begin{tabular}{lccc}
\hline
\hline
Host Galaxy &  &  NGC~3541 \\ 
Redshift &  &  $0.02093 \pm 0.00003$\footnote{\cite{Springob14}}\\  
Distance &  &  $83.90 $~Mpc\\ 
Distance Modulus, $\mu$ &  &  $34.62 \pm 0.54$~mag\\ 
$\textrm{RA}_{\textrm{SN}}$ &  &   $11^{\textrm{h}}08^{\textrm{m}}32.80^{\textrm{s}}$\\
$\textrm{Dec}_{\textrm{SN}}$ &  &  $-10^{\circ}28^{\prime}54.4^{\prime \prime}$\\
$E(B-V)_{\textrm{MW}}$ &  &  0.065 $\pm$ 0.001 ~mag\footnote{\cite{schlegel98,Schlafly+11}}\\
Time of First $o-$band Peak (MJD) &  &  58788.65 $\pm$ 0.01\\
$m_{o}^{\mathrm{peak}}$ &  &  $ 17.34 \pm 0.04$~mag\\
$M_{o}^{\mathrm{peak}}$ &  &  $ -17.43 \pm 0.54$~mag\\
\hline
\end{tabular}
\end{center}
\label{table:Observations}
\end{table}

\section{Optical Observations} \label{sec:obs}

We first obtained publicly available photometric measurements from \ztf{} in the $g$ and $r$ filters, initially published by \citep{Sollerman+20}, extending nearly 200~days after the first reported detection on 2019 October 29.1 (MJD 58785.53).
We also downloaded additional \ztf{} images from the NASA/IPAC Infrared Science Archive\footnote{\url{https://irsa.ipac.caltech.edu/Missions/ztf.html}} to search for detections not reported publically by the automated pipeline. We obtained three epochs of additional photometry where \tsf{} is detected. We isolate the flux of the SN from that of its host galaxy by performing difference imaging using a pre-explosion \ztf{} template image with {\tt HOTPANTS} \citep{Becker15}. Magnitudes were then estimated by modeling each image's point-spread function (PSF) using field stars and subtracting the model PSF from the target. The magnitudes were then calibrated to AB magnitudes from the PS1/$3\pi$ catalog \citep{Chambers16}. These new photometric data are listed in Table~\ref{tab:photometry}.

\begin{deluxetable}{ccc}
\tablecaption{\ztf{} observations of \tsf{}.}
\tablehead{
\colhead{MJD} & \colhead{Mag} & \colhead{Filter}
}
\startdata
58785.53 & $17.43 \pm 0.04$ & r \\ 
58787.53 & $17.52 \pm 0.05$ & r \\ 
58789.53 & $17.49 \pm 0.11$ & r \\ 
\enddata
\tablecomments{Optical photometry of \ztf{} images not reported publicly by the automated pipeline. The magnitudes reported here are the instrumental AB magnitudes without any extinction correction or K-correction applied. \label{tab:photometry}}
\end{deluxetable}

\begin{figure*}[t]
\centering
\includegraphics[width=0.98\textwidth]{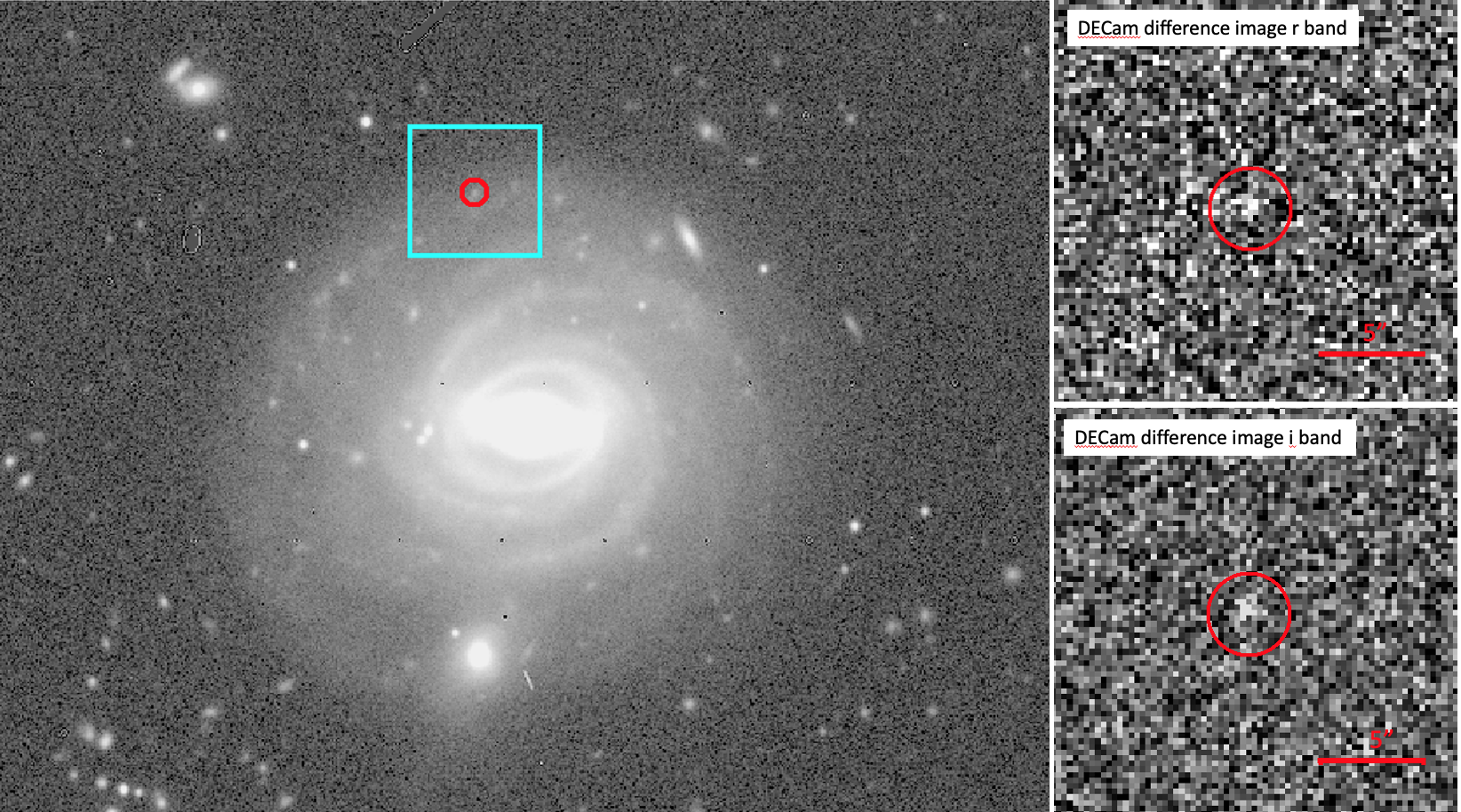}
\caption{
The left panel shows the \decam{} {\it i}-band image taken on 2022 January 12 of \tsf{} in M100 (NGC~3541) at phase +322 days. The position of \tsf{} is marked with a red circle (2" radius). The right top and bottom panels show the different images of that same date in {\it r}- and {\it i}-bands, respectively. The size of the different image cutouts is indicated in the left panel with a cyan box. The SN is detected in the {\it i}-band difference image (S/N $\approx$ 5) and marginally in the {\it r} band. \label{fig:Decam}}
\end{figure*}

\begin{figure*}
\centering
\includegraphics[width=0.98\textwidth]{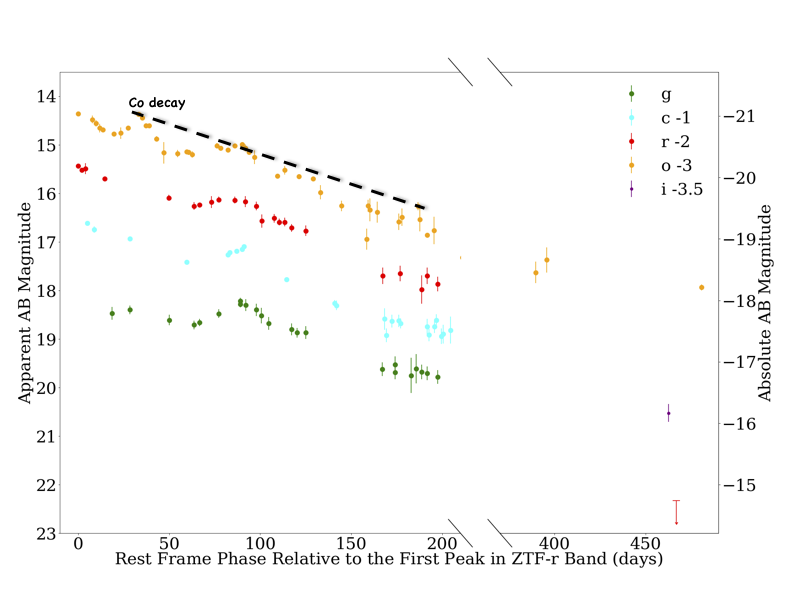}
\caption{Three-peaked light curve of \tsf{} during the first 200 days after the first \ztf{}-\textit{r} bands peak ($\rm t_{\textrm{peak}_r} = 58785.53$). In addition, the late bands' curve evolution until 433 days have been taken by \decam{} in the i- and r-band. That was also the first detection date in all bands. Light curves in \atlas{}-\textit{c,o} band and \ztf{} \textit{g} and \textit{r} bands are also included. The dashed line is the estimate \coVI{} decay.}
\label{fig:lc1}
\end{figure*}

\tsf{} was also observed by \atlas{}, a twin $0.5\,{\rm m}$ telescope system installed on Haleakala and Mauna Loa in the Hawaiian Islands, in the cyan (\textit{c}) and orange (\textit{o}) filters \citep{tonry18}. The \atlas{} images are processed as described in \cite{tonry18} and then photometrically and astrometrically calibrated using the RefCat2 catalog \citep{tonry18ref}. Template generation, image subtraction procedures, and photometric measurements are carried out following \citet{smith2020}. We obtain forced photometry using the ATLAS forced photometry server \citep{Shingles21}. The forced photometry light curve is then cleaned up, and the average flux for each night is calculated using the tools described in \citet[][see Figure~\ref{fig:lc1}(a)]{guevel17, Rest21, RestSofia+23_ATLASClc}. We obtained additional late-time, ground-based imaging of \tsf{} on 2022 January 12, more than 300~days after the explosion, in $r-$ and $i-$band (see Figure~\ref{fig:Decam}) with \decam{} through the \decam{} Extension of the Young Supernova Experiment \citep{DECAT}.

We show the light curves in Figure~\ref{fig:lc1}, where phase $t = 0$~days is defined as 2019 October 29.1 (MJD 58785.53), the date of the first detection and $r$-band peak at $m_r = 17.43\pm0.04$ mag. In the \atlas{}-$o$ band, the earliest detection and first peak occurs 2019 November 1.64 (MJD 58788.64) at $17.34\pm0.04$ mag, $\sim$3 days after first detection in $r$. The last pre-explosion non-detection in \atlas{}-$o$ was obtained on 2019 June 8.27 (MJD 58642.27) below $20.43\ {\rm mag}$. We correct all photometry for Galactic extinction using $R_V = 3.1$ and $E(B-V) = 0.065$ mag, according to the dust maps from \cite{Schlafly+11}, and use the \citet{Barbary16} implementation of the \citet{Cardelli89} extinction law.
An additional cosmological K-correction of $+2.5\log_{10}\left(1 + z\right)$ is also applied. 
After applying these corrections, we measure an absolute magnitude of $M_o = -17.43$ mag at the time of the $o-$band peak. 



\begin{figure*}
\centering
\includegraphics[width=0.98\textwidth]{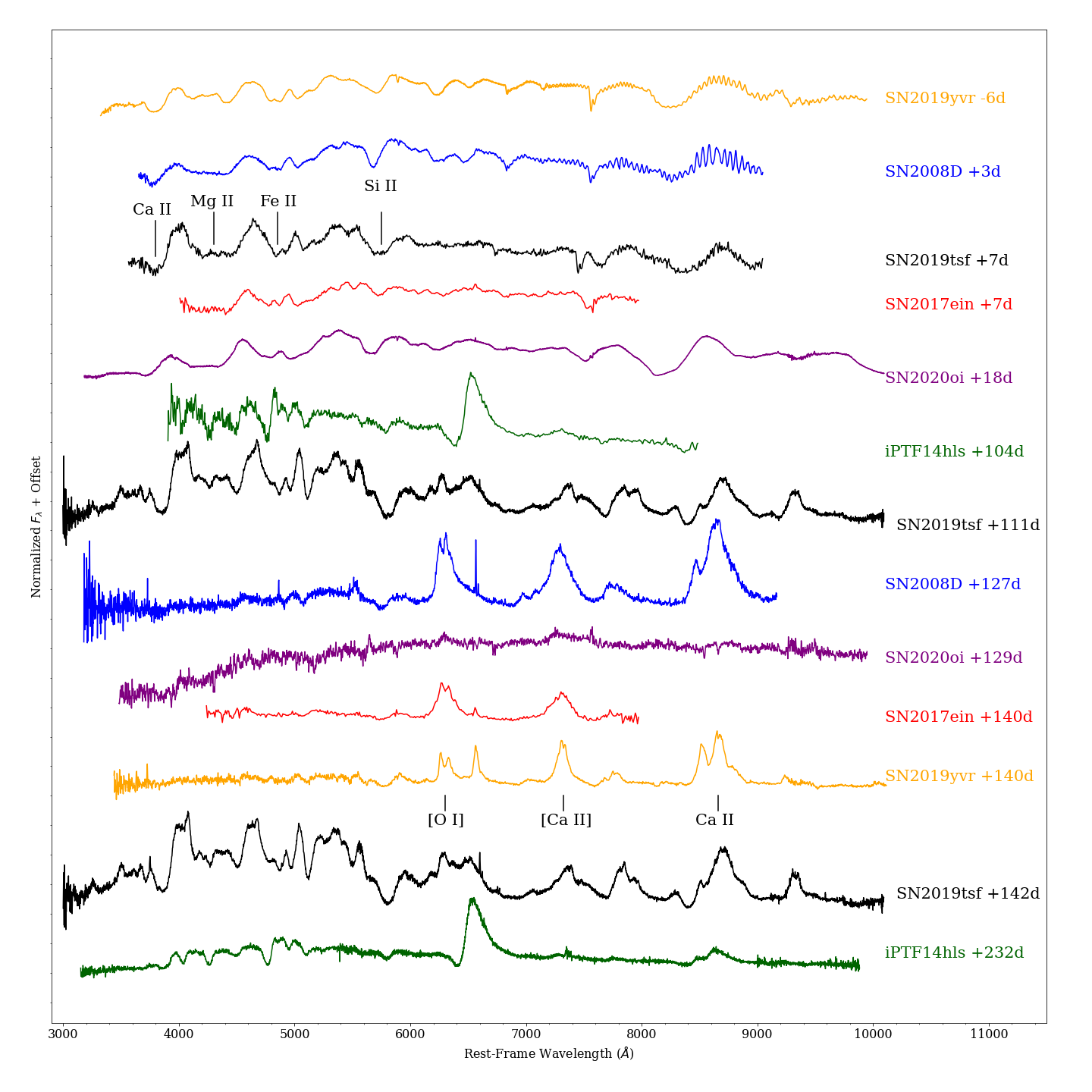}
\caption{The spectra of \tsf{} (in black) taken on +7, +111, and +142 days past the first peak in \ztf-$r$ band from \cite{Sollerman+20}. The later spectrum, taken right after the third peak, is still similar to the typical Type Ib SNe spectra around the peak and shows slow evolution during this period compared to the spectrum taken on +7 days. Spectra of other SNe Ib/c are included for comparison, including SN Ib 2008D (in blue, \citealt{2008D, UCBSample_2008D}), SN Ib 2019yvr (in orange, \citealt{2019yvr}, Auchettl et al in prep.), SN Ic 2020oi (in purple, \citealt{Gagliano+22}), SN Ic 2017ein (in red, \citealt{2017ein}) and abnormal 5-peak \hls{} (in green, \citealt{Arcavi14hls}). Phases relative to their peak in $r$(SN~2020oi, SN~2019yvr), $r$(SN~2008D) or $R$(\hls{}, SN~2017ein) bands are marked near the spectra. The early-time spectra of all these SNe Ib/c are similar to the \tsf{}, but around $\sim 100$ days after the peak, they all evolved dramatically. The time measurements were done from $r$-band peak.} \label{fig:spectra1}
\end{figure*}

We obtained two optical spectra with the ESO Faint Object Spectrograph and Camera \citep[EFOSC2;][]{EFOSC2} on the ESO New Technology Telescope (as part of the ePESSTO survey, \citealt{Smartt15AA}) on 2019 November 5.35 (MJD 58792.35) and 2020 February 13.25 (MJD 58892.25), roughly $\sim 4$ and 100 days after the first peak in the rest frame, as shown in Figure~\ref{fig:spectra1}. Additionally, in figure \ref{fig:spectra1}, we show the \tsf{} spectra at 7 and 111 days. In the early spectra we find evidence of \ion{Ca}{2}, \ion{Mg}{2}, \ion{Fe}{2}, and \ion{Si}{2} features similar to typical, early Ib SNe SN~2008D and SN~2019yvr. \tsf{} hardly evolves in the later phase even until $\gtrsim 100$ days after the peak, and there is no sign of other dominant lines (\ion{O}{1} 6300, \ion{Ca}{2} 7326, and \ion{Ca}{2} 8662) in SNe Ib/c around the same phase, neither are there any signs of hydrogen features.

We also present new late-time comparison spectra of SN~2020oi and SN~2019yvr from the Shane telescope. The spectra were reduced using standard IRAF/PYRAF \footnote{IRAF was distributed by the National Optical Astronomy Observatory, which was managed by the Association of Universities for Research in Astronomy (AURA) under a cooperative agreement with the National Science Foundation} and Python routines for bias/overscan subtractions and flat fielding. The wavelength solution was derived using arc lamps, while the final flux calibration and telluric lines removal were performed using spectrophotometric standard star spectra.

\begin{figure*}
\centering
\includegraphics[width=0.98\textwidth]{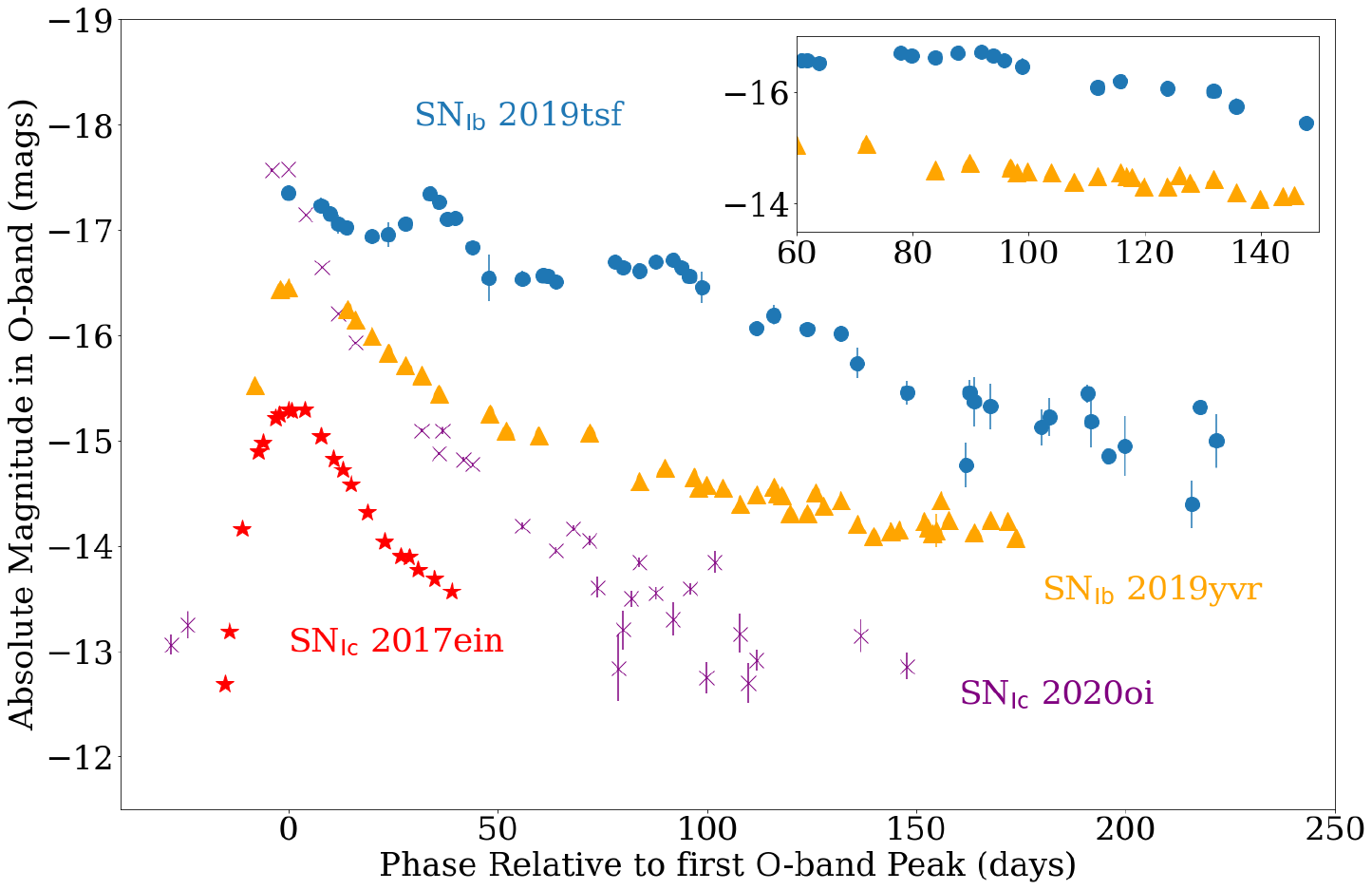}
\caption{Absolute \textit{o}-band light curves constructed from the \atlas{} sample of SESNe Ib, Ic, IIb. This sample includes only \textbf{the two} SNe associated with SNe Ib closest to \tsf{}. Yellow triangles designate the Ib SN~2019yvr, blue dots designate the \tsf{}, red stars designate the Ic SN~2017ein, and purple cross designate the Ic SN~2020oi. \tsf{} behaves differently from the other SESNe, at least until 70 days (relative to the first peak). The late-time evolution, in contrast, bears more similarities to other SNe. The inset plot presents the late-time evolution of two SNe Ib by \atlas{}. They have a similar drop rate in luminosity. The \textit{o}-band points in the inset have been shifted by $\rm \sim +1 mag$ for both Ib SNe. No other shift has been made in the mag axis for any of the other light curves in the figure.} \label{fig:ATLAS_SNe}
\end{figure*}


\section{Radio observations and modeling}   \label{sec:radio_obs}
\subsection{Karl G. Jansky Very Large Array Observations} \label{subsec:radio_obs}

\begin{figure}
\centering
\includegraphics[width=0.58\textwidth]{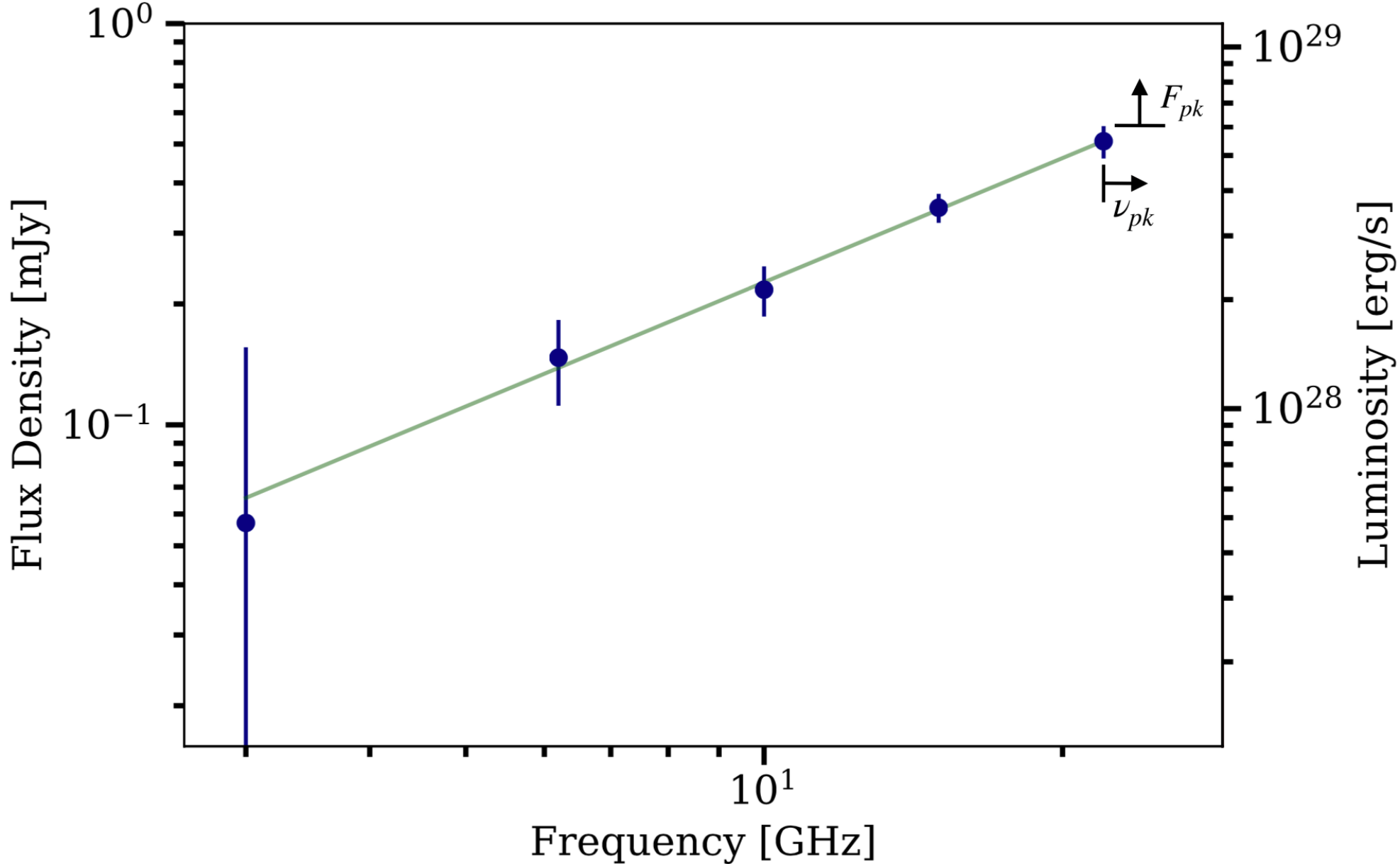}
\caption{VLA observations of \tsf{} (navy points). The green line represents a power-law spectrum $F_{\nu}\propto \nu^{\alpha}$ with a best-fitting index $\alpha = 1.03 \pm 0.05$. The VLA observations capture only the optically thick portion of the spectrum, suggesting that the peak of the radio SED, that we identify with $\nu_{\rm{sa}}$, is above the spectral regime of our observations, or $\nu_{\rm{sa}}>$21 GHz.}
\label{fig:radio}
\end{figure}

We observed the field of \tsf{} with the Karl G. Jansky Very Large Array (VLA) beginning on 2022 January 21 (MJD 59612.29) at a phase of $t = 826.76$~days 
as part of program  VLA/21A-239 (PI  DeMarchi). Observations were taken at S-, C-, X-, Ku-, and K-band, utilizing the WIDAR correlator for maximum sensitivity. The data were reduced using the VLA pipeline in the Common Astronomy Software Applications package (CASA, \citealt{McMullin07}) pipeline version 2020.1.0.36 (CASA version 6.1.2.7) followed by manual inspection, flagging, and reprocessing through the imaging pipeline. For the data taken at C, X, and Ku band we used the image product produced by the pipeline and fit the source using the CASA task \textsc{imfit}. At the S-band, where the source was faintest and corrupting field sources were more prevalent, we applied phase and amplitude self-calibration with a scan-based solution interval to improve the image noise at the target location, which revealed a marginal detection of the target. Our observations are detailed in Table \ref{tab:RadioTable} and plotted in Figure \ref{fig:radio}.

\subsection{Radio Modeling}   \label{subsec:radio_modeling}
In an SN explosion, optical observations sample the slowly expanding ejecta ($v$$\leq$$10^4\,\rm{km\,s^{-1}}$) emitting thermal radiation, while radio observations measure radio synchrotron emission from the fastest ejecta ($v$$\gtrsim$$0.1c$). Radio synchrotron emission originates from the interaction of the fastest SN ejecta with the local CSM, shaped by the mass loss of the progenitor star before the explosion. For the typical radio SN, the result is a bell-shaped spectral energy distribution (SED), with the spectral peak $\nu_{\rm{pk}}$ cascading to lower radio frequencies with time as the blastwave expands and the ejecta become optically thin to synchrotron self-absorption (SSA) and free-free absorption (FFA). In this case, $\nu_{\rm{pk}}$ =corresponds to the SSA frequency $\nu_{\rm{sa}}$. Under these assumptions, by monitoring $\nu_{\rm{sa}}(t)$ and the peak flux density $F_{\rm{pk}}(t)$, we can directly constrain the forward-shock radius $R(t)$ and the post-shock magnetic field $B(t)$, from which  the pre-shock CSM density $\rho_{\rm{CSM}}$ and mass-loss rate $\dot M$ can be derived \citep[e.g.,][]{chev98,chevfrans17}. 

Here, we use the equations presented in \cite{DeMarchi22}, which were derived from \cite{chev98}. In our calculations, we assume that the fraction of total blastwave internal energy $\rho_{\rm{CSM}}v_{\rm{sh}}^2$ (where $v_{\rm{sh}}\equiv dR/dt$, the forward shock velocity) imparted to relativistic electrons is $\epsilon_e=0.1$ and that the post-shock magnetic energy fraction is $\epsilon_B=0.01$. We assume a power-law evolution of the forward-shock radius $R$ as a function of time $t$, such that $R \propto t^q$. However, because the optically thin portion of the spectrum is not observed, we cannot measure $R(t)$ directly. Instead, we adopt $q=0.88$ as in \cite{chev82} for the case of a stripped-envelope SN shock launched by a compact massive star interacting with a CSM wind-density profile.
We are thus able to obtain constraints on the CSM density
around the explosion, littered by the mass-loss history of the progenitor star in the centuries before the core collapse.

For \tsf{}, our multi-frequency VLA data at $t = 827~$--$840$~days sample the optically thick part of the SED and are best fitted with a power-law spectrum $F_{\nu}\propto \nu^{\alpha}$ with index $\alpha = 1.03 \pm 0.05$, suggesting that $\nu_{\rm{sa}}$ is above the spectral regime of our observations, or  $\nu_{\rm{sa}}\gtrsim$21 GHz. Similarly, $F_{\rm{pk}}\gtrsim$ 564 $\mu Jy$. These parameters imply a radius of the emitting region $R\lesssim 10^{15}$cm, which is difficult to reconcile with the forward shock radius of SN at $t>800\,$d after the explosion (for which we would expect $\sim$10$^{17}$~cm).  

Possible explanations fall into two broad categories, both sharing the requirement that the emitting region is \emph{not} a spherically symmetric forward shock launched at the time of the explosion.  The first possibility is that the radio-emitting region consists of a localized overdense ``knot'' of CSM (as was proposed for SN~1986J, e.g., \citealt{Bietenholz1986J, Bietenholz1986J2}), or a disk of material— 
The second possibility includes the emergence of radiation from a newly-formed compact object, for example, in the form of a pulsar wind nebula \citep[PWN; see review by][]{Slane2017, Dong&Hallinan22}.
While we leave a detailed study of the radio emission from \tsf{} (and its temporal evolution) to future work, here we note that the radio observations are consistent with a disk-like geometry of the CSM. 

\begin{deluxetable*}{ccccccc}
\tablecaption{Radio Observations of \tsf{}. }
\tablehead{
\colhead{Start Date} & \colhead{Centroid MJD} & \colhead{Phase$^{\rm{a}}$} & \colhead{Frequency} & \colhead{Bandwidth (GHz)} & \colhead{Flux Density$^{\rm{b}}$} & \colhead{Facility}\\
(dd/mm/yy) &  & (d) & (GHz) & (GHz) & ($\mu$Jy) &
}
\startdata
Jan 21, 2022 & 59600.32 & 811.68 & 10.0 & 4 & 217 $\pm$ 15 & VLA \\ 
Feb 1, 2022 & 59611.30 & 822.66 & 6.2 & 4 & 147 $\pm$ 12 & VLA \\
Feb 1, 2022 & 59611.31 & 822.67 & 22.0 & 8 & 508 $\pm$ 56 & VLA \\
Feb 2, 2022 & 59612.33 & 823.69 & 15.0 & 6 & 347 $\pm$ 23 & VLA \\
Feb 2, 2022 & 59612.29 & 823.65 & 3.0 & 2 &  57 $\pm$ 13  & VLA \\
\enddata
\tablecomments{account force MJD 58788.64, using the central time of the exposure on the source. $^{\rm{b}}$Uncertainties are quoted at 1$\sigma$, and upper-limits are quoted at $3\sigma$. The reported errors account for a systematic uncertainty of 10\% for data at 22\, GHz and 5\%  for all the other frequencies.
\label{tab:RadioTable}}
\end{deluxetable*}


\section{Modeling the Optical Light Curve} \label{sec:lightcurve}

The optical light curves of SN\,2019tsf are exceptionally long-lived for an SN~Ib, lasting more than 450~days after the first detection, and display at least three distinct peaks (Figure~\ref{fig:lc1}). While \atlas{} and \ztf{} missed the pre-explosion phase and early rise \tsf{} (\citealp{Sollerman+20}; c.f., \citealp{Taddia+15}), the first peak, at $M_r = XX$~mag appears to occur around the time of, or just before, the first detection. The second peak, seen prominently as an $0.87$~mag rise in the ATLAS $o$-band light curve, occurs around 40--50 days later. The third peak occurs across all the observed optical bands at a phase of $\approx$90~days. This unique set of light curve properties is difficult to explain with any standard explosion models for SNe~Ib. Moreover, the mixing of \niVI{} with the outermost CSM layers makes it hard to confidently rule out the presence of helium lines, one of the critical indicators of an SN~Ib. Because of these factors make modeling and analyzing \tsf{} more challenging than for a typical SN~Ib. In this section, we explore several progenitor models to attempt to explain the observed properties of \tsf{}.

\subsection{MOSFiT Modeling}

An alternative mechanism for powering the light curves of SESNe is via sustained energy injection from a central compact object formed during the explosion. One often discussed possibility is a rapidly-spinning neutron star (NS) with a strong surface dipole magnetic field strength $B_{\rm NS} \sim 10^{13}-10^{15}\,{\rm G}$, which spins down and releases its rotational energy in the form of a magnetized wind on a timescale of days to years \citep{Ostriker&Gunn1971, Kasen&Bildsten10, Woosley2010}. This wind inflates behind the supernova ejecta a compact magnetized nebula of charged particles (e.g., \citealt{Metzger+14}) that powers the optical emission through thermalization by the ejecta of the high-energy synchrotron and inverse-Compton radiation released by these particles \citep{Vurm&Metzger21}. Though mainly considered as a model for Type~I SLSNe (e.g., \citealt{MarguttiMetzger17, Nicholl+17,moriya+22}), magnetar engines have been invoked to boost the luminosities of other types of SN or SN-like transients (e.g., \citealt{Yu+13, Sukhbold&Thompson17, Prentice+18, Gomez+19}).

We first attempt to model the optical light curves using {\tt MOSFiT} \citep{Guillochon+18}, a Python code designed to fit the light curves of transients for a variety of power sources using {\tt emcee} MCMC package \citep{foreman13}. We model the light curve with a magnetar central engine to derive physical parameters and estimate the explosion time, which was missed by the photometry of both \atlas{} and \ztf{}. For a complete description of the {\tt MOSFiT} implementation of the magnetar central engine {\tt slsn} model, see \cite{Nicholl+17}. In our models, we use the same model priors as the ones in \cite{Nicholl+17}, with the exception that we allow for a slower spin period extending to $40$ ms to accommodate the lower luminosity of \tsf{}, compared to the SLSNe modeled in \cite{Nicholl+17}. We run the model using 150 walkers for $\sim$15,000 steps and test for convergence by ensuring the models reach a potential scale reduction factor $<1.3$ \citep{gelman92}.

\begin{figure}
\includegraphics[width=0.52\textwidth]{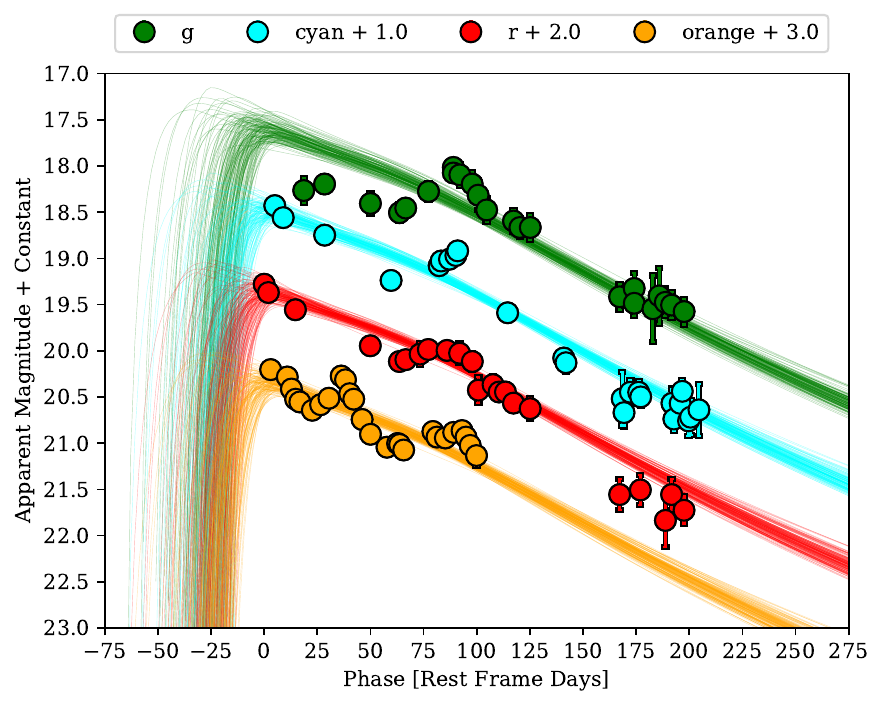}
\caption{Optical light curves of \tsf{} modeled with a magnetar central engine model using {\tt MOSFiT}.}
\label{fig:mosfit}
\end{figure}

The resulting {\tt MOSFiT} light-curve models are shown in Figure~\ref{fig:mosfit}, and the posterior distributions of the main parameters are shown in Figure~\ref{fig:corner}. We find a magnetic field $B = (2.3^{+2.1}_{-0.7}) \times 10^{14}$ G, a spin period of $p_{\rm spin} = 22 \pm 3$ ms, an optical opacity $\kappa = 0.14 \pm 0.05\ {\rm cm}^2{\rm g}^{-1}$, a gamma ray opacity of $\log(\kappa_\gamma / {\rm cm}^2 {\rm g}^{-1}) = -0.54^{+0.21}_{-0.27}$, an ejecta mass of $M_{\rm ej} = 0.86^{+0.67}_{-0.46}$ M$_\odot$, and an ejecta velocity of $v_{\rm ej} = 7000 \pm 2600 $ km s$^{-1}$. {\tt MOSFiT} only allows us to model the light curves with a simplified one-zone model, and it is, therefore, unable to replicate the multi-peak structure of the light curve. Nevertheless, we can recover approximate parameters from a model that reproduces the overall shape and duration of the light curve. From these models, we also measure a total radiated energy of $E_{\rm rad} \sim 3 \times 10^{49}\,{\rm erg}$, integrated up to a phase of 250 days.

\begin{figure}
\centering
\includegraphics[width=0.50\textwidth]{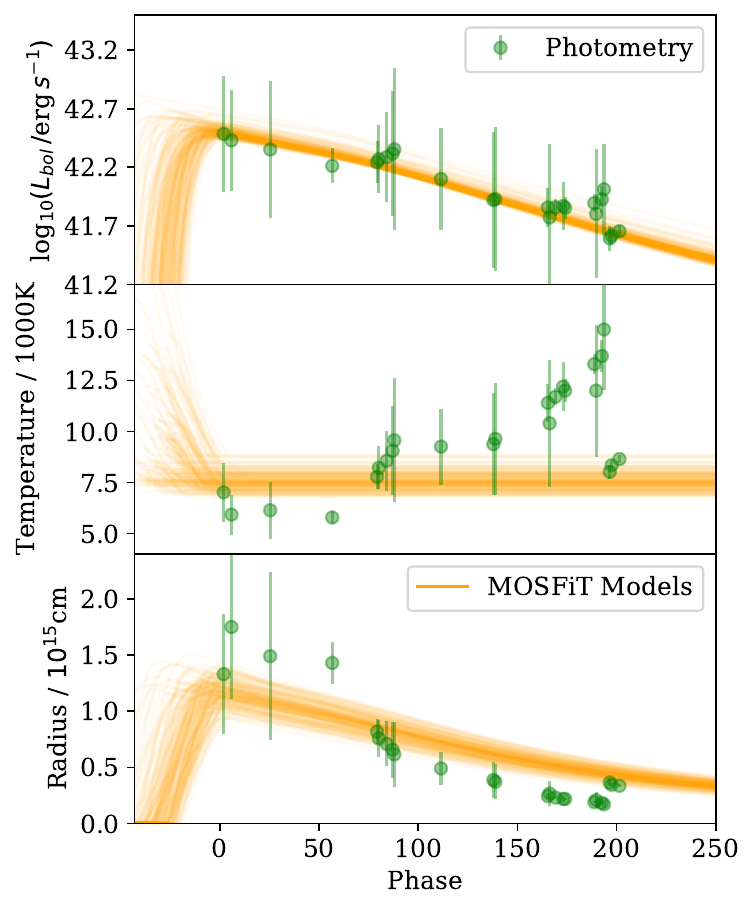}
\caption{Bolometric luminosity (\textit{top}), temperature (\textit{middle}), and radius (\textit{bottom}) evolution of \tsf{}. The green points are measurements estimated from the photometry using {\tt Superbol}, while the orange lines are the equivalent measurements obtained from {\tt MOSFiT}.}   \label{fig:superbol}
\end{figure}

Additionally, we measure the bolometric luminosity, blackbody radius, and temperature evolution of \tsf{} using the {\tt Superbol} code \citep{Nicholl18_superbol}. {\tt Superbol} works by first interpolating the light curves of all individual bands using a polynomial function to account for the times of non-concurrent photometry. 
{\tt Superbol} extrapolates the blackbody function to account for missing UV and IR flux outside the observed bands. The final bolometric luminosity, radius, and temperature estimated from the photometry are shown in Figure~\ref{fig:superbol}. We see that the measurements from {\tt Superbol} largely match the equivalent values measured from the {\tt MOSFiT} model, with the exception that {\tt Superbol} estimates a steep rise in temperature after $\sim$150 days. It is hard to determine if this temperature rise is real given that the peak of the blackbody function is well into the UV, and we lack photometry bluer than $g$-band. This effect is reflected in the large error bars of those temperature measurements.

\subsection{Magnetar model + Radioactive decay + CSM interaction} \label{subsec:modelingMagnetar}

Models for SN light curves have also been proposed that invoke a combination of a magnetar engine, ejecta-CSM interaction, and $^{56}$Ni decay (e.g., \citealt{Chatzopoulos+13, ChenZ+22}). For \tsf, we propose that such a combination of power sources may be necessary to explain the unusual multi-peaked light curve.

The rotation energy of a NS born with a rotational period $P = 2\pi/\Omega$ and angular velocity $\Omega$ can be written $E_{\rm rot}= I_{\rm NS}\Omega^2/2$, where $I_{\rm NS} \approx 1.6\times 10^{45}$ g cm$^{2}$ is the moment of inertia of a 1.4\,${\rm M}_{\odot}$ NS \citep{Lattimer&Schutz05}. 
The magnetar loses rotational energy according to
\be
    - \frac{dE_{\rm rot}}{dt}  = {I_{\rm NS}} \Omega \dot{\Omega} = L_{\rm sd} ,  \label{Edot_rot}
\ee
where the dipole spin-down luminosity can be written \citep{Contopoulos+99, Metzger+15}
\be
    L_{\rm sd}  = \frac{B_{\rm NS}^{2} R_{\rm NS}^{6}\Omega^4}{c^3}  ,  \label{Lsd}
\ee
and $R_{\rm NS} \approx 12$ km is the NS radius.  The solution to Eq.~(\ref{Edot_rot}) is given by
\be
    L_{\rm sd} = - \frac{dE_{\rm rot}}{dt}  = \frac{E_{\rm rot,0}}{t_{\rm sd}}\left(1+\frac{t}{t_{\rm sd}}\right)^{-2} ,  \label{Ldot_rot}
\ee
where the characteristic dipole spin-down time is
\be
    t_{\rm sd}  = \frac{E_{\rm rot,0}}{L_{\rm sd,0}} = \frac{6 c^3 I_{\rm NS} }{B^{2}\Omega_0^2 R^6_{\rm NS}} ,  \label{tsd}
\ee
and the subscript `0' denotes evaluation at $t = 0$.

Optical radiation escapes, and the supernova light curve peaks on the radiative diffusion timescale (e.g., \citealt{Kasen&Bildsten10}),
\be
t_{\rm pk} \sim t_{\rm d} \sim \left(\frac{3\kappa_i M_{\rm ej}}{4\pi v_{\rm ej}t}\right)^{1/2},
\ee
where $M_{\rm ej}$ and $v_{\rm ej}$ are the mass and mean velocity of the SN ejecta and $\kappa_{i}$ is the optical Rosseland mean opacity.  Insofar that $t_{\rm d} \gtrsim t_{\rm sd}$, a large portion of the magnetar's rotational energy $E_{\rm rot,0}$ can be transferred to the supernova ejecta and in this limit we have
\be
\frac{1}{2}M_{\rm ej}v_{\rm ej}^{2} \simeq E_{\rm rot,0}  \label{Mej}
\ee
The Rosseland mean opacity $\kappa_{i}$ depends on the composition and interior temperature $T$ of the ejecta following \citep[see their Fig.~3]{Kleiser&Kasen14}. The photosphere temperature at $t_{\rm pk}$ can be written,
\be
T_{\rm eff} = \left(\frac{L_{\rm pk}}{4\pi \sigma R_{\rm pk}^{2}}\right)^{1/4}
\ee
where $R_{\rm pk} = v_{\rm ej}t_{\rm pk}.$

\begin{figure*}
\centering
\includegraphics[width=0.98\textwidth]{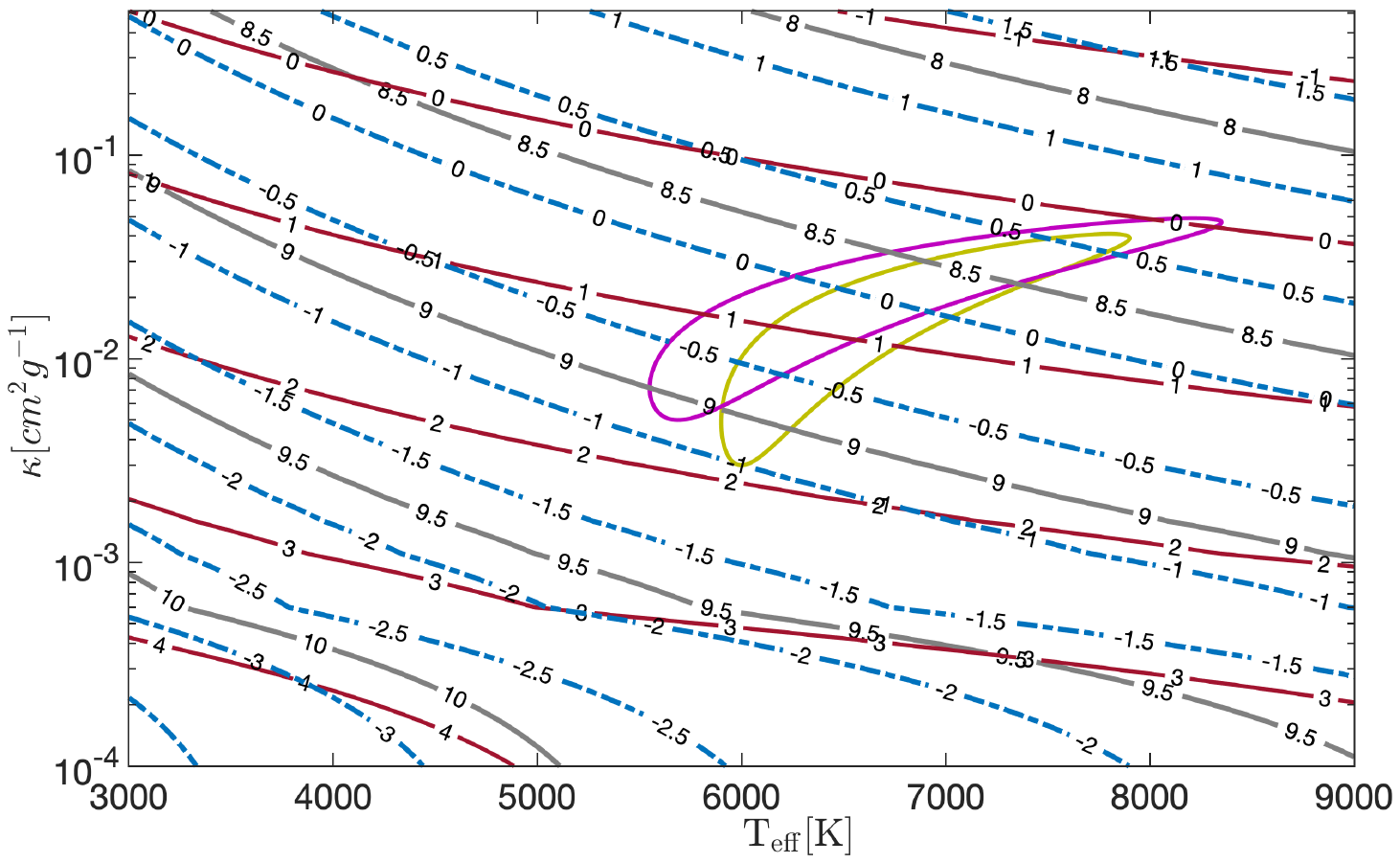}
\caption{Map of ejecta temperature $T$ versus optical opacity $\kappa_i$ overlaid with contours of the magnetar surface field strength $\log_{10} (B_{\rm NS}/10^{14}{\rm G})$ (red solid lines) and supernova ejecta mass $\log_{10}(M_{\rm ej})$ (blue dashed lines), for assumed magnetar spin period $P_0 = 3$ ms (chosen post-hoc because it gives values $M_{\rm ej} \sim 1-7M_{\odot}$ following progenitor expectations as ejecta mass, velocity, etc.). The gray dash line is the ejecta velocity ($0.005-0.04 c$, speed of light) derived from the condition $t_{\rm pk} \lesssim t_{\rm sd}$.
The purple and yellow ellipses are the 1$\sigma$ contours for $L_{\rm pk}$ and $t_{\rm pk}$ based on the multi SE \tsf{} light curve data. The opacity range we find agrees with that found by \citet{Kleiser&Kasen14} for hydrogen-deficient ejecta in this temperature range $T \rightarrow T_{\rm eff}$.}
\label{fig:B14Mej_v2}
\end{figure*}

Combining the above equations, we can use the observed properties of the first ($L_{\rm pk} \sim 2.5 \pm 0.1 \times 10^{42}{\rm erg\,s^{-1}},t_{\rm pk} \sim {\rm 0 d}$) and second light curve peaks ($L_{\rm pk} \sim 7 \pm 0.08 \times 10^{42}{\rm erg\,s^{-1}},t_{\rm pk} \sim {\rm 50 d}$) to create contours of $B_{\rm NS}$ and $M_{\rm ej}$ in the space of $T_{\rm eff}$ and $\kappa_{i}$, as shown in Fig.~\ref{fig:B14Mej_v2}. Here, we assume $P_0 = 3$ ms.  We see that the first/second peak can be explained by a magnetic field of $B = \{5,70\}\times 10^{13}\,{\rm G}$ and $M_{\rm ej} = \{0.75,7.6\}~\Msun$ respectively. We further see that the $\kappa_i(T)$ in the range of our solutions agree with those for He-rich ejecta found by \citet[][their Fig.~3]{Kleiser&Kasen14}. In Figure~\ref{fig:mosfit}, we include those values to fit the light curve with the magnetar engine using \texttt{MOSFiT}.

The magenta engine can potentially explain the first peak in the light curve of \tsf{}; however, as mentioned above, a combination of power sources, including radioactive heating and eject-CSM interaction, is likely necessary to explain the subsequent peaks. SN ejecta's opacity strongly depends on its ionization state and temperature. Again, based on the ejecta-mass and magnetic-field contours in the temperature--opacity phase space shown in Figure~\ref{fig:B14Mej_v2}, the maximum ejecta mass of \tsf is $M_{\rm ej} = 2.63\,\Msun$, with $B_{\rm NS}= 3.54\times 10^{14}\,{\rm G}$, $\kappa = 0.04 \ {\rm cm}^{2}\ {\rm g}^{-1}$, $T\sim 8300\,{\rm K}$ (constrained based on Figure~3 of \citealt{Kleiser&Kasen14}). Accordingly, we conclude that the \tsf{} first peak may be adequately explained with a predominant magnetar engine without additional contributions of the radioactive heating or the CSM interaction.

Still, the magnetar-only model for SN Type Ib cannot reproduce the multi-peaked light curve of \tsf{}. We thus suggest that the second peak at 40--50 days may be attributed to radioactive heating, while the third peak at $\approx$90~days may be explained by weak CSM interaction. Assuming that the second peak is caused directly by radioactive decay, the observed decline rate must be slower than or close to the \coVI{} decline $\rm \sim 0.01\ mag\ day^{-1}$ (see Figure~\ref{fig:lc1}). Following the third peak, the light curves decline to $\approx$200~days fade at $\approx$0.0182\ mag\ day$^{-1}$, more rapidly than the \coVI{} decay rate. Therefore, this rate is inconsistent with pure \coVI{} decay. The second peak (yellow ellipse in Figure~\ref{fig:B14Mej_v2}) also lies in the regime of a typical SNe~Ib with a magnetar engine and may thus be explained by an additional contribution from radioactive heating. One possibility to explain the third peak is a powerful CSM interaction with asymmetric geometry, which we also explore in more detail in \citep{SmithHB17,Andrews&Smith18,Brethauer+22,Thomas+22}. 

\subsection{A warped disk in a triple system} \label{subsec:modelingtriple}

\tsf\ is a very rare and possibly unique Type~Ib SN. As discussed in Section~\ref{sec:discussion}, given that no other Type~Ib SN has been observed to have three peaks, a \tsf-like SN likely happens between once per few $10$ and once per few $100$ Type Ib SNe. Furthermore, while \tsf{} shows no signatures of hydrogen in its spectrum, similarly to typical Type Ib SNe, the significant amount of disk-like CSM inferred from the radio observations (Section~\ref{sec:radio_obs}) requires an explanation. In this section, we describe an option relatively common sequence of events that may occur in Type Ib SN progenitor systems such that the resulting SN bears similarity to \tsf{}.

Firstly, we point out that typical SN Ib progenitors will likely have a bound tertiary companion at the moment of explosion. The majority of stars more massive than $8$~--~$10\,{\rm M}_\odot$ are born in binaries or higher-order multiples \citep{Sana2012,Sana2013}. \citet{Moe2017} compiled diverse observations of the diversity of early-type stars and estimated that $8\,{\rm M}_\odot$ stars have on average $1.3\pm0.2$ companions with masses of at least $0.8\,{\rm M}_\odot$. Similarly, they found that $12\,{\rm M}_\odot$ stars have on average $1.6\pm0.2\,{\rm M}_\odot$ companions. We also note some uncertainty regarding the exact number of triple and higher-order multiple companions in the literature. For example, \citet{Abt1990} found that $8\,{\rm M}_\odot$ stars have on average $2.0$ companions with masses above $0.8\,{\rm M}_\odot$ and \citet{Rizzuto2013} inferred that $12\,{\rm M}_\odot$ stars have on average $1.3$ companions with masses above $1.2\,{\rm M}_\odot$. We can conservatively say that $f_{\rm 3rd}=10$~--~$100\,\%$ of all SNe Type Ib progenitors have triple or higher-order multiple companions.

The orbital separations of tertiary companions are approximately log-uniformly spaced from $3$ times the semi-major axis of the inner binary to $\approx$5$\times 10^3\,{\rm AU}$ \citep{Moe2017}. Assuming that in typical a SN Ib, the primary is stripped by stable mass transfer onto the secondary companion with an orbital separation between 30 and 800\,${\rm R}_\odot$ \citep[e.g.,][]{Marchant2021}, then the outer tertiary companion will have the following distribution of orbital separations: $f_{\rm a 3rd} \equiv \dfrac{\mathrm{d} N_{\rm 3rd}}{\mathrm{d} \log_{10} (a_{\rm 3rd}/{\rm AU})} = \dfrac{1}{\log_{10} a_{\rm 3rd, max} - \log_{10} a_{\rm 3rd, min}} = 0.24$~--~$0.38/{\rm dex}$.

Recently, \citet{Laplace2020,Laplace+21} suggested that some fraction of SN Ib progenitors may be undergoing mass transfer at the moment of the explosion. Studies, including \citet{Akashi+15, Pejcha2017, Decin2020}, have also explored the role of mass transfer and associated mass loss in binary systems. These investigations examined the formation of disk-like outflows, jets, and conical outflows as mechanisms for mass loss in binary systems undergoing mass transfer. Depending on the evolutionary stage of the star during the interaction, some such systems may produce He-rich CSM and result in SNe~Ibn \citep{Laplace2020,Laplace+21}. These occur at a rate compared to SNe~Ib of approximately $f_{\rm Ibn}=0.1$. 
As \tsf{} did not exhibit H lines at any phase, we assume it was produced in a similar scenario.

The CSM of mass-transferring systems involving tertiary companions will be perturbed. The most studied type of such distortions is disc warping \citep[e.g.,][]{stefano10,2021MNRASSokeBearParasiteCESN,2021MNRASGlanzTCE}. This effect has been observed and modeled, in particular, in protoplanetary discs, \citep[e.g.,][]{Marino2015, Andrews2020}, accretion discs around spinning black holes \citep{Bardeen1975}, galaxies \citep{Binney1992} including the Milky Way \citep{Momany2006} and its Galactic center \citep{Bartko2009}, as well as AGN disks \citep[e.g.,][]{Nayakshin2005}. Likewise, warped discs have been thoroughly modeled both analytically and numerically across these respective fields \citep[e.g.,][]{TremaineDavis, Lai2014, Nealon2018}. Disc warping from interactions with a tertiary companion typically leads to an inclined secondary disk and may range from mild warps/twists to breaks or gaps. A broken disk could naturally explain the presence of a third light-curve peak as the explosion propagates into this two-component medium. 

The evolution of warped and broken disks is complicated, and full hydrodynamic simulations are usually needed to reproduce it in a self-consistent way. 
The morphologies of warped disks are driven by the external torques, the inner total angular momentum of the disk, and the radial pressure gradient \citep{Miller&Krolik13, Salcedo+18}. The balance between viscous and external torques governs whether breaks in such disks can occur. 

A lower limit on the inner radius of the outer disk is set by the separation of the tertiary from the center, which can be calculated from the stability criteria for a triple system \citep{2001AsresthMardlingTripleStability}:
\begin{align}
    R_{\rm warp} \gtrsim a_{\rm out} \gtrsim C(1+M_3/M_{\rm bin})^{2/5}
\end{align}
where $C\approx 2.8$ and $a_{\rm out}$ is the tertiary separation.

The maximal location of the disk break could be estimated by comparing the viscous torque to the external one (as in \citealp{Nixon2013}):

\begin{align}
R_{\rm warp}\lesssim \left(\frac{1}{4}\mu |\sin 2\theta|\left(\frac{H}{R_{\rm disk}}\right)^{-1}\alpha^{-1}\right)^{1/2}a_{\rm out}
\end{align}

\noindent
where $\mu \equiv M_3/(M_{\rm bin}+M_3)$ is the reduced mass of the triple, $\theta$ is the inclination angle of the disk, $H/R_{\rm disk}$ is the aspect ratio of the disk, and $\alpha$ is the Shakura-Sunyaev parameter \citep{ShakuraSunyaev73}. 
Hence, given the disk's parameters and the components of its progenitor system, the location of the break can be estimated. This, in turn, constrains the location and time of the induced SN peak.

\begin{figure}[t]
    \centering
    \includegraphics[width=\linewidth]{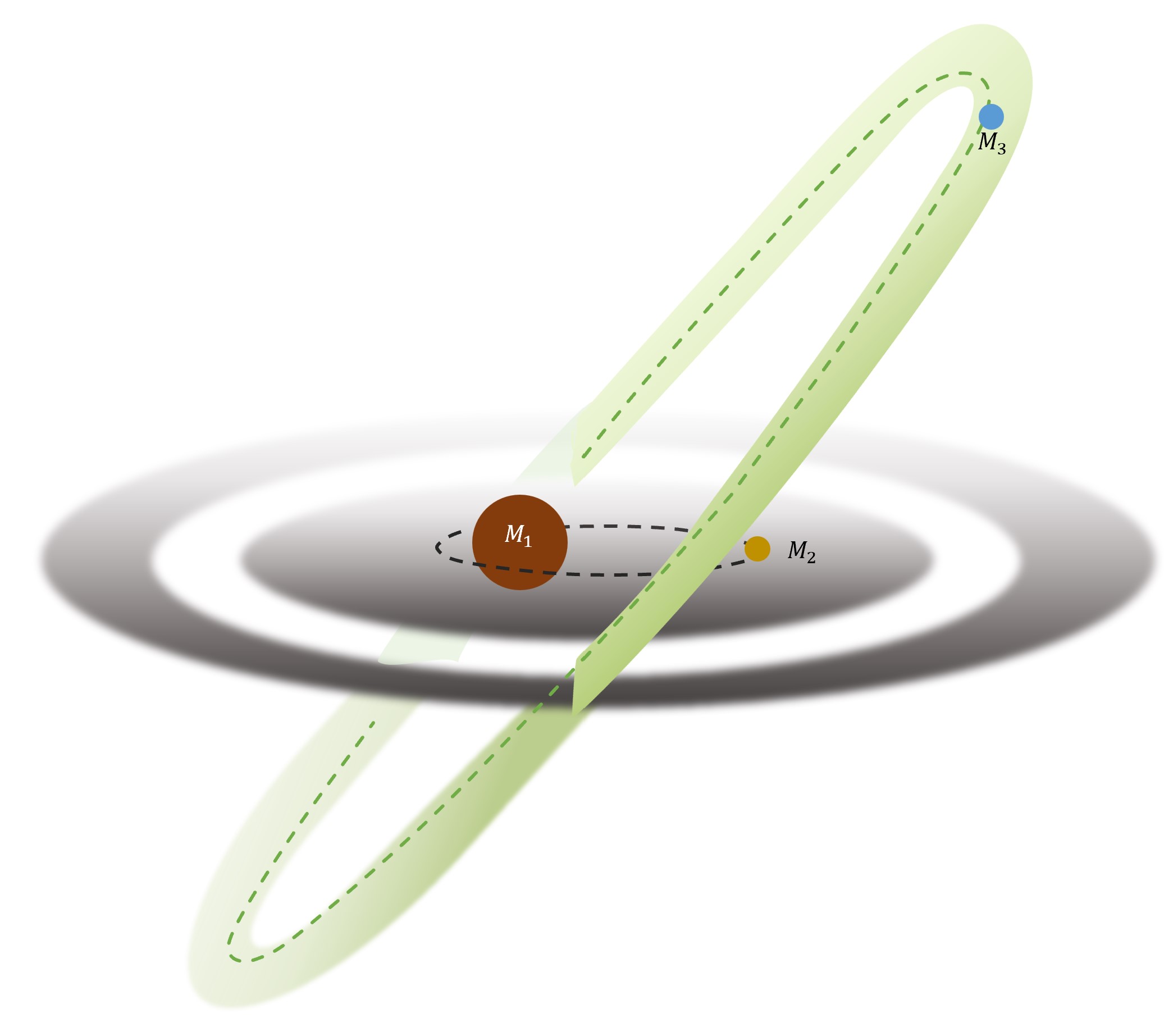}
    \caption{Initial condition of the triple system described in Sec. \ref{subsec:modelingtriple}, demonstrating the warped circumbinary accretion disk, with the secondary inclined disk that has formed from the interaction of the tertiary component with the circumbinary disk, leading to the later disk break.} 
    \label{fig:TripleStep1}
\end{figure}

\begin{figure}
    \centering
    \includegraphics[width=0.95\linewidth]{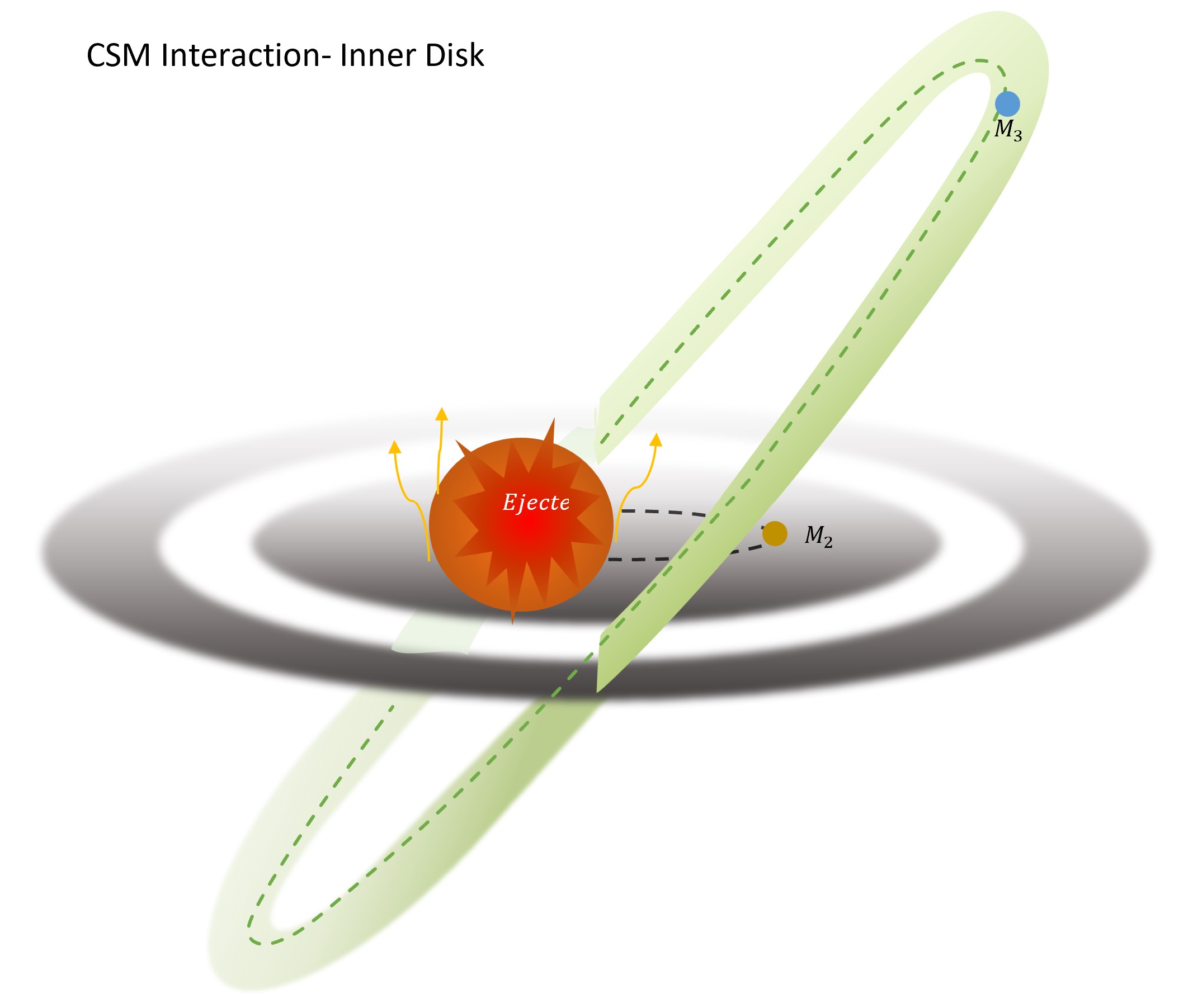}
    \includegraphics[width=0.95\linewidth]{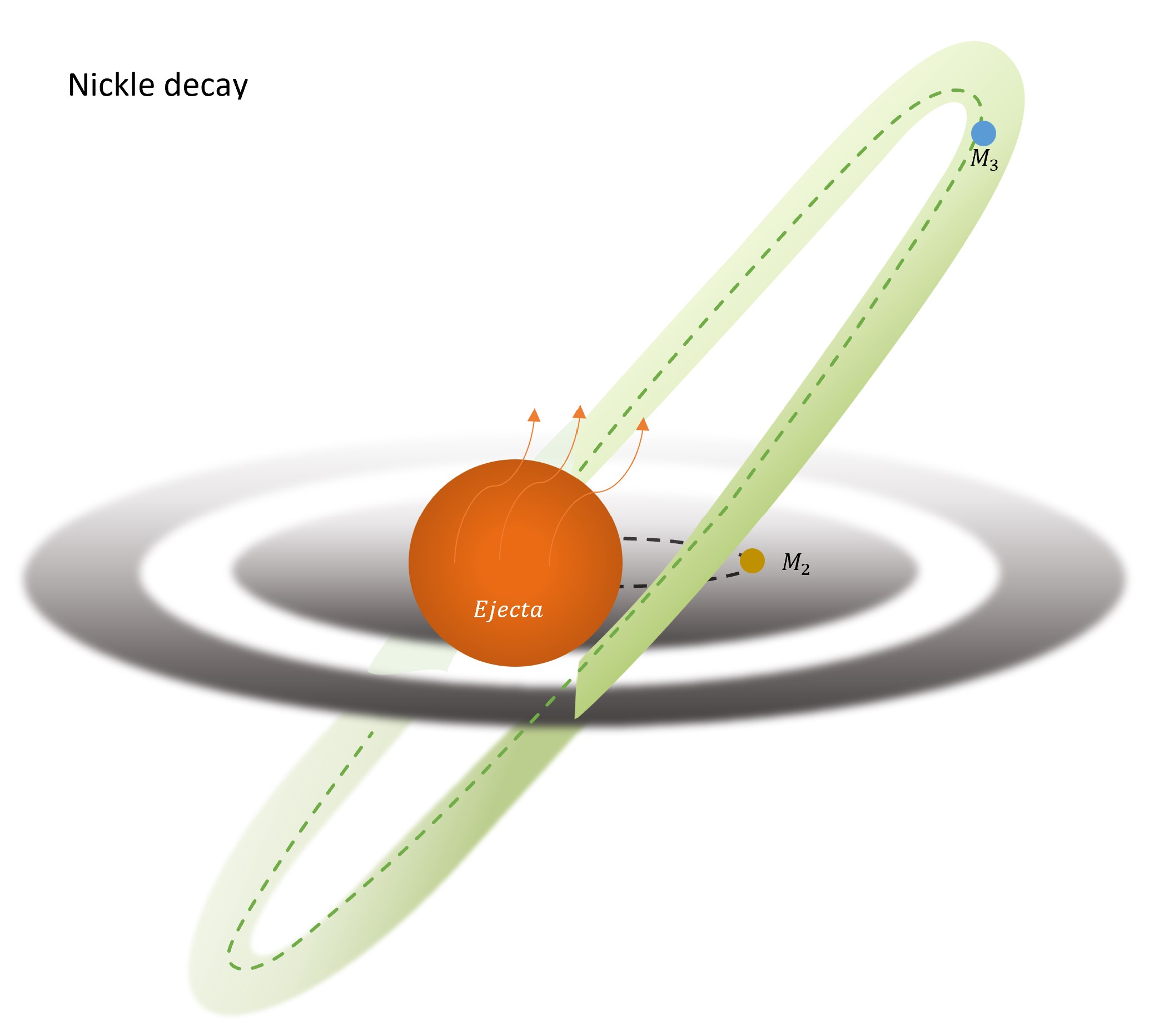}
    \includegraphics[width=0.95\linewidth]{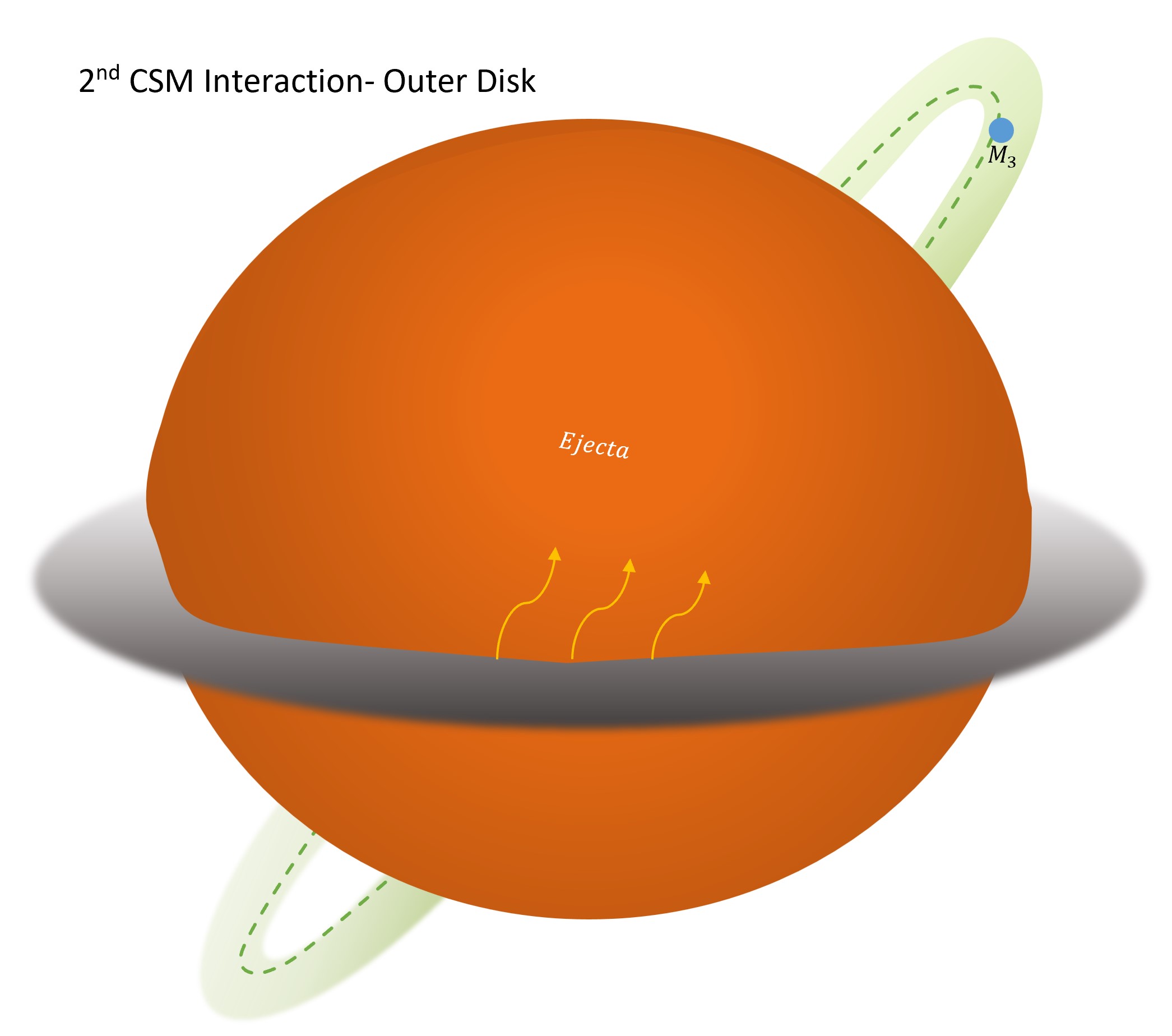}
    \caption{The three events (from top to bottom) leading to the SN peaks in the triple system model described in Sec. \ref{subsec:modelingtriple}. The interaction with the inner disk forms the first peak, the nickel decay of the explosion causes the second, and the third peak is caused by the interaction of the ejecta with the outer disk. The SN is observed edge-on to both planes, so hydrogen lines can hardly be observed.}
    \label{fig:TriplePeaks}
\end{figure}

Assuming \tsf{}\ belongs to the subclass of SNe Ib from mass-transferring systems with a tertiary companion, we illustrate the suggested formation channel in Figs. \ref{fig:TripleStep1} and \ref{fig:TriplePeaks}. In this scenario, the supernova interaction with the inner disk generates the first peak, and the supernova interaction with the outer disk generates the third peak. In contrast, the nickel decay from the supernova produces the second peak. The timing of the third peak may be estimated as $t_{\rm 3rd peak}=R_{\rm break}/v_{\rm SN, shock}$. A $1\,{\rm dex}$ range of the third peak times, e.g. $50$~--~$500\,{\rm d}$, thus corresponds to tertiary companions within $1\,{\rm dex}$ range of separations, e.g. $200$~--~$2000\,{\rm AU}$, assuming $v_{\rm shock}=7000\,{\rm km}/{\rm s}$. Overall, the fraction of SNe Type Ib that evolves in this way is $f_{\rm 3-peak} = f_{\rm Ibn} \cdot f_{\rm 3rd} \cdot f_{\rm a, 3rd} \cdot one \,{\rm dex} = 2.4\cdot 10^{-3}$~--~$3.8\cdot 10^{-2}$, or once per $26$--$420$ SNe Ib and, more specifically, once per $2.6$--$42$ SNe Ibn. As discussed in Section~\ref{sec:discussion}, this conservative estimate of the occurrence rate of \tsf{}-like events agrees with current observational constraints.

\begin{figure*}
    \centering
    \includegraphics[width=0.99\textwidth]{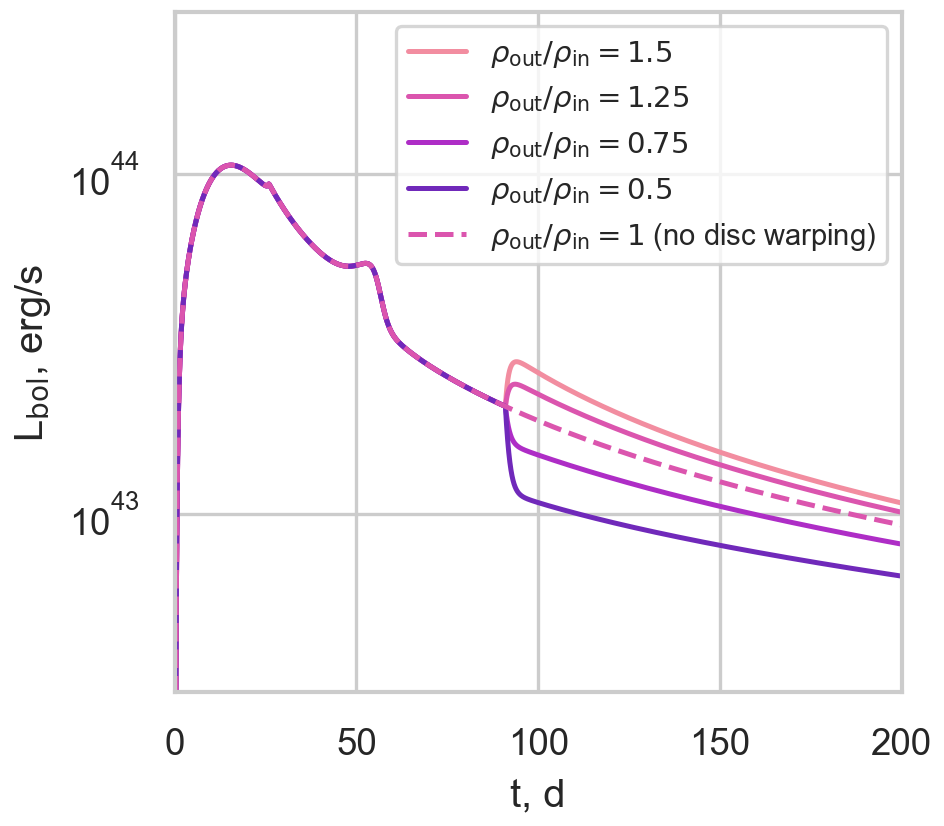}
    \caption{Synthetic optical Bolometric lightcurves for the disc-ejecta interaction model \citep{Metzger&Pejcha17} for \tsf{} for the case when a triple companion introduces a local discontinuity in the disc profile. These profiles qualitatively reproduce multi-peaked structures similar to those observed in \tsf{}.} 
    \label{fig:DDModel}
\end{figure*}

We illustrate the qualitative behavior of the light curves corresponding to our scenario by modifying the model for ejecta-disk interactions by \citet{Metzger&Pejcha17}. In this model, the collision of supernova ejecta with the disc produces a shock, and the radiation from the shock subsequently diffuses outwards through the supernova ejecta. We consider a conservative case in which the binary loses mass through the L2/L3 point into a disc with an opening angle of $0.2$ at a constant rate of $10^{-7}\,{\rm M}_\odot/{\rm yr}$ (e.g., \citealp{Brethauer2022, JG2022}) and expansion velocity of $50\,{\rm km} {\rm s}^{-1}$. The disc density profile has a local increase or decrease at $100\,{\rm AU}$ separation, corresponding to 

We show the results of these simulations in Figure~\ref{fig:DDModel} for a range of local density increases and decreases by up to a factor of two and for a case of no disc discontinuity. A discontinuity in the radial disc profile modifies the light curve by producing an additional bump feature similar in timing and magnitude to the one observed in \tsf{}. 

It should be noted that the model by \citet{Metzger&Pejcha17} has some limitations, including in the assumed hydrogen-based opacities or the lack of complete radiative transfer through the supernova ejecta. Therefore, these light curves should only be seen as a demonstration that disc discontinuities may introduce additional peaks in supernova light curves similar to the one observed in \tsf{}.

In summary, the warped/broken disk scenario is a natural consequence of the idea that mass-transferring binaries produce SNe Type Ibn, as proposed in \citet{Laplace2020}, and that a significant fraction of stars are observed in triples or higher-order multiples. This model is qualitatively consistent with the observed shapes of the bolometric light curve and is generally consistent with the CSM morphology inferred from the radio observations (see Section~\ref{subsec:radio_modeling}). It is also consistent within an order of magnitude with the observed rates (see further discussion in Section~\ref{subsec:Uniqueness}). 

On the other hand, drawing a conclusive connection between this scenario and \tsf{} would require more detailed light curves, spectral modeling, and detailed population synthesis for the progenitor systems.


\section{Discussion} \label{sec:discussion}

\subsection{SN \hls{}-like Events As Closely Related Type of Transients} 
\label{subsec:SN2014hls}

\begin{figure}
\centering
\includegraphics[width=0.49\textwidth]{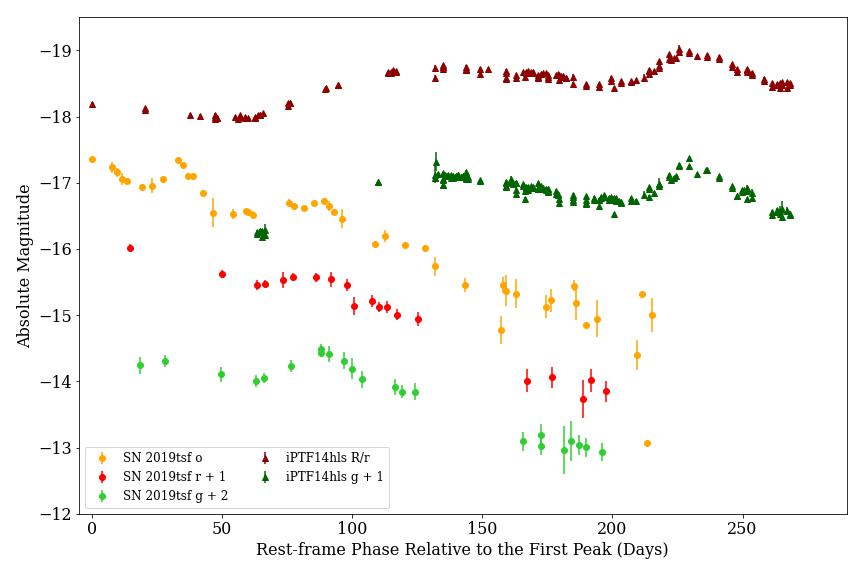}
\caption{The comparison between the absolute magnitude of \tsf{} and \hls{} in multiple bands. Notice that the wavelength range of ATLAS-$o$ band $5582-8249$\AA\, largely overlaps with $R-$ and $r-$bands. Additionally, \hls{} has two extra peaks after $\sim 300$ days after the first peak, while \tsf{} has already faded around the same phase.}
\label{fig:lc_14hls_comp}
\end{figure}

The three peaks in the \tsf{} light curve require profound analysis to understand their similarity to the \hls{} light curve. Undeniably, the multi-peaked light curve of \hls{}, which has not been observed in any other SN II, has a total integrated luminosity of $2.2 \times 10^{50} \rm erg$ during its first 600 days of evolution \citep{Arcavi14hls, Sollerman+19}. The remarkable light curve of \hls{} may be attributed to the interaction with a dense CSM, as suggested by \citet{Andrews&Smith18, WangL+22}, based on the emergence of spectroscopic interaction signatures at late phases. 
In the case of \tsf{}, there are no spectra available after $\sim 220$~days to compare with those of \hls{} when the signs of CSM interaction became evident (see figure \ref{fig:lc_14hls_comp}).
We note that various models have been proposed to explain the peculiar behavior of \hls{}, some indicating the presence of more than power source, including CSM interaction, pulsational pair-instability supernova (PPISN), magnetar formation, and a standard envelope interaction system involving a NS and a RG star with jets \citep[see][]{Woosley+07Natur, Andrews&Smith18, SokerGilkis18, Woosley18, Modjaz+19, KaplanSoker20, WangL+22}.

At a phase of $t \sim 220\,{\rm d}$ after the explosion of \tsf{}, the radius of the ejecta is estimated to be $r \sim 10^{15}\,{\rm cm}$ and the bolometric luminosity is measured to be approximately $L_{\rm bol}\sim 10^{41.7}\,{\rm erg}~{\rm s}^{-1}$  (Figure~\ref{fig:superbol}).
The shock velocity, the speed at which the supernova shock wave is propagating through the ejecta, can be estimated as $v_{\rm s} \sim (2rL_{\rm bol}/M_{\rm ejecta})^{1/3}\sim 10^{18.8}\cdot M_{\rm ejecta}^{-1/3}$. Using this relation and adopting typical ejecta masses for SNe Ib, we obtain $v_{\rm s} \sim 8\times 10^2$--$3\times 10^3\,{\rm km}~{\rm s}^{-1}$. By assuming $v_{\rm s} \equiv v_{\rm ejecta}$ and using equation \ref{Mej}, $v_{\rm s} \sim 10^{-4.7}t_{\rm d}^{-1/6}(c/\kappa)^{-1/12} \sim 0.003-0.008c$. 

Additionally, the spectra of \tsf{} show no evidence of hydrogen during its first, second, and third peaks. This lack of hydrogen is a hint that \tsf{} may have a different nature compared to \hls{}, which has been proposed to be associated with a PPISN.

Theoretical studies of PPISNe predict that the ejecta would clear out most of the hydrogen envelope, as described by\citep[see, e.g.,][]{Woosley+07Nat, Woosley18, ChenJung21}.
The lack of hydrogen in the spectra of \tsf\ makes it unlikely to be a PPISN. Furthermore, the kinetic energy associated with the pulse of a PPISN is typically greater than $\gtrsim 4\times 10^{51}\ \rm erg$, which is higher than the estimated kinetic energy for \tsf{}.

A crucial feature of PPISNe that is hard to reconcile with the features of \hls{} is the $H_\alpha$ velocity of $\sim 8000\ \rm km\ s^{-1}$, which is not available for \tsf{} due to the lack of spectroscopy study. On the other hand, the ejecta velocity of \tsf{} from our {\tt{MOSFiT}} modeling is $v_{\rm ej} = 7000 \pm 2600 $ km s$^{-1}$, which is very compatible with \hls{}. 

In PPISNe, CSM interaction between pre-supernova shells can lead to luminosities $L\simeq 0.5\dot{M} v_s^3 / v_{\rm wind}\simeq few \times 10^{42}\,{\rm erg}\cdot {\rm s}^{-1}$, potentially extending beyond the optical window. 
Furthermore, PPISN models typically require kinetic energy higher than $4\times 10^{50}\,{\rm erg}$ during the intervals between pulses. This energy budget is significantly larger than what is available for \tsf{}, suggesting that an alternative mechanism may be responsible for its powering. One such alternative is the central engine model, which involves the spindown of a millisecond magnetar. Detailed discussions and investigations regarding this possibility can be found in sub-sections \ref{subsec:modelingMagnetar} and \ref{subsec:radio_modeling}.

\subsection{Modeling Caveats}

Since the \atlas{} light curves of \tsf{} do not contain the early rising phase of the explosion and only include two bands (o and c), our modeling fit to the observations is relatively unrestricted. In particular, our results are sensitive to our assumptions about the explosion time and initial conditions for the SN (i.e., initial binary system mass, disk expansion velocity, and rotational period). We, therefore, preferred to adopt realistic initial values for masses, mass-loss rates, and opacity based on the observations of other SNe~Ib.
Most such SNe have properties consistent with the models we use to explain the properties of \tsf{}.One such fundamental property is the estimated ejecta mass ($\rm M_{ej} \sim 1.7-5.2 \Msun$) of ordinary SNe~Ib \citep{Drout+11, Dessart+16}, which agrees well with our modeling results $\rm M_{ej} = 2.63M_\odot$ (see sec. \ref{subsec:modelingMagnetar}).

We note that the slow spectroscopic evolution of \tsf{} at late times shares similarities with strongly CSM-interacting supernovae, particularly SNe IIn and IIn/Ia. In these types of supernovae, the spectra often exhibit relatively narrow hydrogen emission lines. These emission lines are attributed to a dense and slowly evolving CSM formed through the mass loss of the progenitor star before the supernova event occurs. \citep[see][]{smith14, SmithHB17, Filippenko1997, Fox+15} as they provide more detailed information and analysis on the spectroscopic properties of CSM-interacting supernovae, including SNe IIn and IIn/Ia.

Therefore, the most significant modeling challenge is to explain the multiple peaks of the light curve of \tsf{}, with limited bands, jointly with the rare H/He-deficient spectrum. At the same time, the observations allow us to infer that the geometry of the CSM interaction is likely non-spherical with a highly asymmetric mass distribution, perhaps in a disk.
Such geometries allow the CSM interaction to hide the H/He under the photosphere \citep{Smith+15, Metzger&Pejcha17, Andrews&Smith18}. Furthermore, the peculiar CSM profile allows for moderate material exchange between the disk and the ejecta (see \citep{SmithHB17, Andrews&Smith18}). Eventually, the ejecta will expand freely and wholly devour the CSM \citep{Smith+15}. In the context of other core-collapse SNe (e.g., SNe Type IIn), the current understanding is that the CSM interaction may be the most plausible pathway for explaining the protracted, re-brightening, or irregular light curves, e.g. \citet{Andrews&Smith18}.

\subsection{How unique is \tsf{}?}
\label{subsec:Uniqueness}
The diversity observed in the light curves and spectra of SESN likely reflects various outcomes resulting from numerous stellar evolution pathways of massive stars. These possible pathways may include stable mass transfer or standard envelope evolution. The fact that the late-time spectrum of \tsf{} has still not entered the nebular phase by 142~days after the peak is a distinctive and unusual feature.

The nebular phase is characterized by the dominance of line emission from the radioactive decay of isotopes, typically iron-group elements, and indicates a later stage in the supernova evolution. The mean free path of photons $l_{\gamma}\sim \left(\kappa \rho \right)^{-1} \lesssim \frac{2}{3}$, where the values of the opacity $\kappa$ and density $\rho$ are taken from \ref{fig:mosfit} and \ref{fig:B14Mej_v2}, 
plays a crucial role 
the transition to the nebular phase. In the case of \tsf{}, the values of these parameters inferred from Fig \ref{fig:superbol} and \ref{fig:B14Mej_v2}, are unusual compared to typical SESNe. This unique feature further emphasizes the need for in-depth observational and theoretical investigations to understand better the underlying physical processes and the nature of the progenitor system of \tsf{}.

The diversity of light curves and spectra of SESNe likely reflects the diversity of the possible outcomes of multiple stellar evolution pathways of massive stars. The SESN progenitor may be stripped through stable mass transfer or common envelope evolution, likely corresponding to SNe Type Ib or SNe Type Ic. The late \tsf{} spectrum time was detected at $142\,{\rm d}$ after peak, at which point the SN still was not in the nebular phase. The average of mean free path of the photons is  $l_{\gamma}\sim \left(\kappa \rho \right)^{-1} \lesssim \frac{2}{3}$, the values of opacity and density taken from \ref{fig:mosfit} and \ref{fig:B14Mej_v2}. This is a distinctive and strange feature of \tsf{} distinguishing it from most other SNe potentially similar to this event.

A stripped core-helium burning star may produce dense CSM through winds or pre-supernova eruptions before exploding \citep{Matsumoto2022}. At low metallicities, the donor may expand significantly due to stellar evolution when entering the shell-helium burning phase and overfill the Roche lobe \citep{Laplace2020}, as likely happened in \tsf{}. Roche lobe overflow necessarily leads to mass ejection (can also be conservative), the geometry of which may be diverse. In the case of low-mass stars, for example, this may include spherical, jetted, and disk-like features \citep{SokerLivio94, Decin2020}. However, at the high mass transfer rates expected from pre-supernova light curves, the mass loss will likely proceed through a disk \citep{Lu2022}. Furthermore, a significant fraction of massive stars are expected to have a tertiary stellar companion \citep{Sana2012}. The stellar companion may be commonly located within a few $1000\,{\rm AU}$ of the supernova, potentially affecting the geometry of mass loss such that the effect is observable in the first $100\,{\rm d}$ of the light curve. Finally, the observed signatures of the asymmetric CSM will depend strongly on the viewing angle.

A detailed calculation of the occurrence rate of multiple peaks in SESN light curves is limited by the uncertainties of various massive stellar evolutionary pathways. It is outside the scope of this study. Still, we can expect that multi-peaked SNe Ibn should be relatively common among SNe Type Ib in low-metallicity environments, assuming the model of \citet{Laplace2020} is correct. Therefore, analyzing the distribution of occurrence times of the third peak could provide a tempting new lens on the tertiary companion properties of massive stars, assuming that more SNe-like \tsf{} are found in the future. These conclusions are in qualitative agreement with the current observations of SNe Type Ib. For example, 224 SNe Type Ib/c are currently listed on the (\ztf{}) Bright Transient Survey catalog 
\citep{Fremling2020, Perley2020}, likely representative of stably stripped stars. Among these, 20 SNe are Type Ibn, potentially indicative of SNe from stably stripped stars in low-metallicity hosts overflowing their Roche lobe. \tsf{} is a unique SN Type Ib that is about $20$ times rarer than SNe Type Ibn and about $200$ times rarer than typical single-peaked SNe Ib. However, the true occurrence rate of multi-peaked SESNe may be a factor of several higher since not all the SNe Type Ib in \ztf{} are sampled well enough to detect multiple peaks. The best example of such missed SNe is \tsf{} itself, which was recognized as a triple-peaked SN only thanks to the \atlas{} data.
Dedicated follow-up observations, including comprehensive photometric and spectroscopic monitoring, are necessary to advance our knowledge of Type Ib SNe and rare events like \tsf{}. By combining these observations with theoretical modeling efforts, we can work towards a more comprehensive understanding of the underlying physics and the diverse range of progenitor scenarios that give rise to events like \tsf{} or \hls{}.

\subsubsection{Implications and Context}
\label{subsub:OtherSESN}

The absence of late-time spectra for \tsf{} prevents confirmation of CSM interaction, though weak H/He features and the lack of a nebular phase suggest either low-mass ejecta or obscured interaction—potentially due to disk-like CSM shaped by binary mass transfer. Similar signatures in other SESNe support a role for binary evolution in producing warped or asymmetric CSM. Homogeneous, multi-wavelength monitoring particularly of early color, mass-loss rates, and late-time emission—is essential for connecting light curve structure to progenitor properties. Further observational campaigns that include late-time spectroscopic observations of \tsf{} and detailed modeling efforts will be instrumental in unraveling the nature of this unique supernova and shedding light on the mechanisms responsible for its distinct spectral properties and behavior. For the complex CSM models, such as the ejecta model proposed in \cite{SmithHB17}. Usually, the prominent peak of the light curve can only last as long as it takes for the heat to diffuse out, which lasts a month or so, depending on the ejecta mass. We observed narrow H, or He lines from the slow CSM ahead of the shock and photoionized by the shock itself. However, there are thought to be cases where the CSM interaction may be strong, but the standard signatures of CSM, like narrow lines and X-rays, may be hidden from view. For more details about ejecta running into disk, see \cite{smith14, Smith+15, Andrews&Smith18}.

\subsubsection{Non-spherical CSM}

Notably, a twisted and non-spherical CSM profile would be primarily the result of binary evolution during which the primary fills its Roche lobe; this occurs during an interaction between stellar winds and mass transfer \citep{yoon17}. In this binary system, the primary star fills its Roche lobe during the evolving stage, depending on the evolutionary state of the stellar core when this occurs and the initial separation. The mass transfer could happen in more than one phase \citep{Lauterborn70, Yoon15}. If the mass ratio of the stellar components is sufficiently large, the system will experience unstable mass transfer, leading to a joint envelope event or binary mergers. In the other case, when the initial mass ratio is mild, the mass transfer is stable, and the primary will end with a small amount of hydrogen (in the envelope). Such a binary stellar evolution scenario could lead to a SN~Ib with a disk surrounding it. As we showed above, the H/He-lines (in the spectra) then can become hidden below the photosphere after that disk is surrounded by the fast SN ejecta \citep{Dessart+12, Smith+15, Andrews&Smith18}. Finally, for the unknown SESN progenitors, multiple empirical observational relations can be used to infer the ejecta properties, such as the progenitors' ejecta mass or explosion energy \citep{Drout+11, Dessart+15, Dessart+16}.

\section{Conclusions} \label{sec:conclusion}

This paper presents a rare SESN Type Ib known as \tsf{} with 430~days of \atlas{} data after the explosion. This transient was also observed four months after the optical peak in the radio with the VLA. This kind of transient shows multiple peaks in its optical light curves that are difficult to connect directly to a specific progenitor scenario. There are few known multi-peaked SNe of different types \citep[see][]{Arcavi14hls, Sollerman+19, Sollerman+20}. Studying such rare and atypical events like \tsf{} provides valuable insights into the diversity of supernova progenitors and the range of physical processes involved in their explosions. Three distinct peaks in the light curve, including a relatively initial solid peak (only apparent in the ATLAS data), a second peak 35 days later, and a third peak around 90 days after the first detection, highlight this SN's complex and unique behavior. The continued absence of explicit hydrogen in the spectrum at $\rm \leq 145d$ days and the lack of clear spectral signatures of CSM interaction, as discussed in \cite{Sollerman+20}, are particularly noteworthy. 
The absence of hydrogen in the spectra suggests that the progenitor star likely had already lost its outer hydrogen envelope before the explosion, corresponding to a Type Ib or Type Ic supernova. Meanwhile, the lack of clear CSM interaction signatures indicates that the interaction with surrounding material may be relatively weak or highly asymmetric.

In Figure~\ref{fig:ATLAS_SNe}, we compare the \tsf{} light curve to some prototypical SNe Ib and Ic such as SN~2019yvr, SN~2017ein, SN~2020oi respectively, and we conclude that the \tsf{} light curve behaves like a SN Ib. Nevertheless, we estimate the $\rm {}^{56}Ni$ mass by following \cite{Dessart+11, Dessart+15, Khatami&Kasen19, Gagliano+22} and find that it could be powered by $0.07-0.12\ \Msun$. 
Additionally, we explore the possibility that a magnetar powers the entire light curve and find the best fit of a magnetar of the magnetic field $B \sim 2 \times 10^{14}$ G and spin period $P_{\rm spin} \sim 22$ ms, and an ejecta mass of $M_{\rm ej} \sim 0.9$ M$_\odot$. One of the conclusions taken from figure \ref{fig:spectra1} is that SESN can lack dominant H/He emission lines or even hide them completely \citep[e.g.][]{Matheson+01, Dessart+11}.
\citet{Andrews&Smith18} explained that such a CSM interaction might not reveal itself in the spectral evolution due to a particular geometry that hides the interaction site. However, the actual modeling of such a mechanism is less explored.

We present alongside our optical data the first epoch of our radio VLA observations (Table \ref{tab:RadioTable}). As only the optically thick spectrum of the SED can be constrained, we are only able to speculate (estimate radius, mass, velocity, etc) that there is a sufficiently dense CSM (either a clump or disk) located at $R<10^{15}\,{\rm cm}$ from the explosion. Ongoing radio observations and analyses will be published in future work.

We have considered several physical scenarios to explain the multi-peak SESN \tsf. These include (i) SN ejecta interaction with a warped disk, (ii) SN ejecta interaction with asymmetric disk-like CSM formed by binary interaction, (iii) a purely magnetar-dominated model, and (iv) the magnetar-model with CSM interaction \ref{subsec:modelingMagnetar}. The ejecta running into a disk scenario, where the H-lines are hidden during the early evolution, is consistent with the asymmetric CSM properties intrinsic to Type-Ib explosions. The extraordinary model here is the warped disk scenario as we emphasize the stellar evolution in \ref{subsec:modelingtriple}. The results of this \tsf{} highlight how necessary both the early-time observations and the late time multi-wavelength observation are for understanding the nature of SN progenitors.  
\tsf{} it part of a broader, overlooked class of explosions where binary interaction and compact CSM structures shape both the light curve and the observed spectra. Future radio monitoring and deep late-time spectroscopy will be essential to confirm this interpretation and to assess how common such progenitor scenarios are among SESN.

\section{Acknowledgements} \label{Sec:ack}

The author thanks Kevin Schlaufman and Arshia Maria Jacob for helpful discussions. A.B. acknowledges support for this project from the
European Union’s Horizon 2020 research and innovation program under grant agreement No. 865932-ERC-SNeX.
MR acknowledges the generous support of the Azrieli fellowship. 
The UCSC team is supported in part by NASA grant 80NSSC20K0953; NSF grants AST-1815935; the Gordon \& Betty Moore Foundation; the Heising-Simons Foundation; and by a fellowship from the David and Lucile Packard Foundation to R.J.F.
This work has made use of data from the Asteroid Terrestrial- impact Last Alert System (\atlas{}) project and the Young Supernova Experiment (YSE; \cite{Jones+21}. The Asteroid Terrestrial- impact Last Alert System (\atlas{}) project is primarily funded to search for near-earth asteroids through NASA grants NN12AR55G, 80NSSC18K0284, and 80NSSC18K1575; byproducts of the NEO search include images and catalogs from the survey area. The ATLAS science products have been made possible through the contributions of the University of Hawaii Institute for Astronomy, the Queen s University Belfast, the Space Telescope Science Institute, the South African Astronomical Observatory, and The Millennium Institute of Astrophysics (MAS), Chile. This work was partially funded by Kepler/K2 grant J1944/80NSSC19K0112 and HST GO-15889, and STFC grants ST/T000198/1 and ST/S006109/1.
This research is also based on data from the Astro Data Archive at NSF’s NOIRLab. These data are associated with observing program(s) 2020A-0415 (PI A. Rest), 2021A-0275 (PI A. Rest), and 2021B-0325 (PI A. Rest). NOIRLab is managed by the Association of Universities for Research in Astronomy (AURA) under a cooperative agreement with the National Science Foundation. Raw imaging data were processed with the \decam{} Community Pipeline (\cite{Valdes+14}, ASPC, 485, 379).
We further acknowledge that the National Radio Astronomy Observatory is a facility of the National Science Foundation operated under a cooperative agreement by Associated Universities, Inc.
The authors wish to thank the referee, whose numerous questions and points led us to add detailed explanations of the striped envelope SNe; furthermore, investigating the overlap with the mysterious multipeak \hls{} is an important goal. We believe this has led to a paper that is much more approachable to readers.

\facilities{\emph{DECam, VLA}}

\software{Astropy \citep{Astropy+18}, \texttt{MOSFiT} \citep{Guillochon+18},  \texttt{Superbol} \citep{Nicholl18_superbol}, \texttt{CASA} \citep{McMullin07}, \texttt{emcee} \citep{foreman-mackey13}, \texttt{HOTPANTS} \citep{Becker15}, and NumPy \citep{2020NumPy-Array}}

\bibliographystyle{aasjournal} 
\bibliography{references}

\clearpage
\appendix

\begin{figure}
\centering
\includegraphics[width=0.98\textwidth]{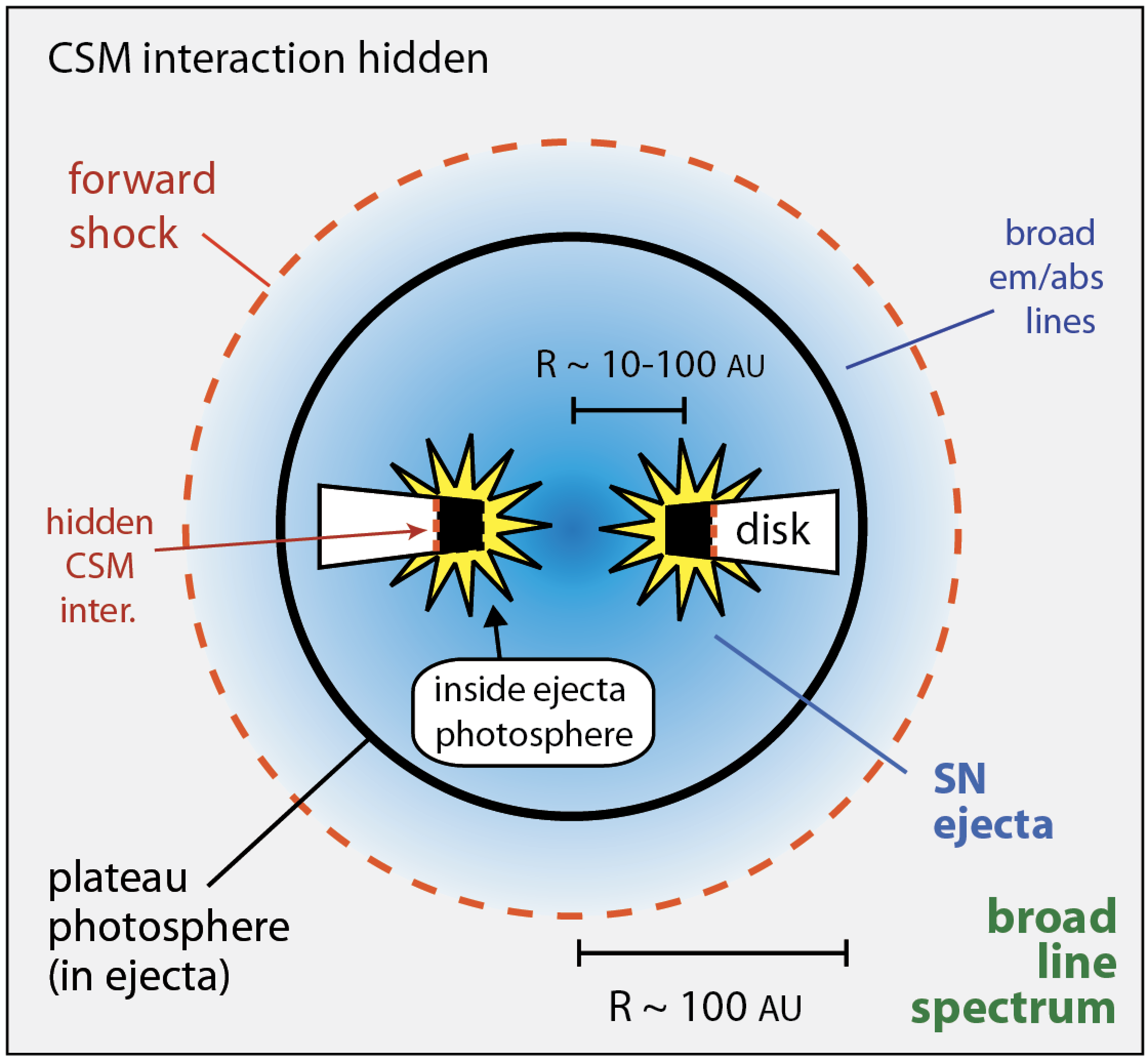}
\caption{A sketch showing how CSM interaction with a compact disk can power the luminosity even if CSM interaction signatures (narrow lines, X-rays, etc.) are not observed. This is adapted from essentially the same sketch shown for the Type IIn event PTF11iqb by \citet{Smith+15}, and then also adopted for the peculiar SN~II event \hls{} by \citet{Andrews&Smith18}. The model for \tsf{} may be nearly identical, except that the SN ejecta are H depleted, making the emergent spectrum a Type Ib, and the radial density distribution of the inner CSM disk may be somewhat different. The key point is that the expanding SN ejecta engulfs the compact disk, so while the radiation from this CSM interaction (represented by yellow spikes) can reheat and ionize the ejecta and continue to power the light curve as long as the disk lasts, the radiation is trapped inside the optically thick SN ejecta and is thermalized before it can escape being seen by an external observer. Thus, an observer sees only broad lines from the fast SN ejecta photosphere. The resulting light curve may appear bumpy if the radial density structure in the disk is not uniform. \label{fig:hiddendisc}}
\end{figure}

A Corner plot of our posterior distributions is provided in 1D and 2D posteriors from the light-curve modeling of the best-fit parameters of \tsf{} with MOSFiT. We used an ensemble-based Markov chain Monte Carlo with 250 walkers to sample our parameter space through 50000 iterations. 2D histograms have been smoothed using a Gaussian kernel with a standard deviation of 1.

\begin{figure}[ht!]
    \centering
    \includegraphics[width=0.98\textwidth]{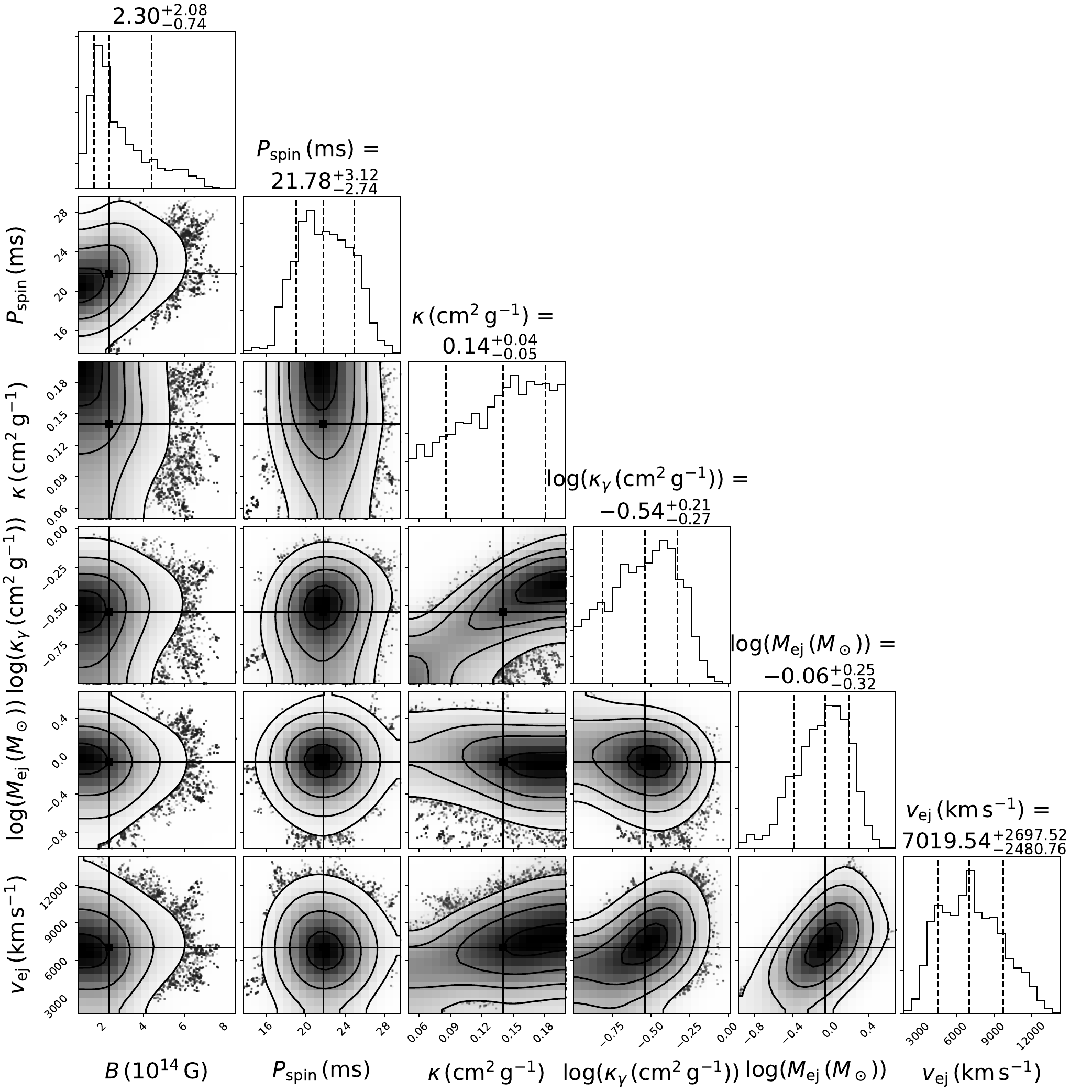}
    \caption{Posterior distributions of the best-fit parameters of the models shown in Figure~\ref{fig:mosfit}. Figure generated using {\tt corner} \citep{Foreman16}.}
    \label{fig:corner}
\end{figure}

\end{document}